\documentclass{article}
\usepackage{amsmath}
\usepackage{graphicx}

\begin{document}

\title{Spontaneous emission of an atom placed near a nanobelt of elliptical
cross-section}
\author{D.V. Guzatov and V.V. Klimov
\thanks{vklim@sci.lebedev.ru} \\
P.N. Lebedev Physical Institute, Russian Academy of Sciences,\\ 53
Leninsky Prospect, 119991 Moscow, Russia}

\date{\today}
\maketitle

\begin{abstract}
Spontaneous emission of an atom (molecule) placed near a nanocylinder of
elliptical cross-section of an arbitrary composition is studied. The
analytical expressions have been obtained for the radiative and nonradiative
channels of spontaneous decay and investigated in details.
\end{abstract}

\section{Introduction}

As was first pointed out by Purcell \cite{ref1} one can change the atomic
decay rate by using a resonant cavity. It is well known now that not only
resonant cavities, but also any material body may exert influence on the
spontaneous decay rate \cite{ref2}. Moreover, control for the spontaneous
decay rate is widely used in practice at elaboration of new efficient light
sources \cite{ref3}.

Due to rapid development of nanotechnologies influence of nanoparticles of
different shapes on radiation of atoms, molecules and nanocrystal quantum
dots is of key importance. This direction often referred to as nanophotonics
\cite{ref4}. Influence of spherical \cite{ref5}-\cite{ref7}, spheroidal \cite%
{ref8}-\cite{ref10} and even ellipsoidal \cite{ref11} nanoparticle on
radiation of atoms and molecules is investigated in details.

Of special interest is the influence of dielectric fibers or metallic wires
of different cross-sections on the decay rate of a single atom. This is
because the charged wires were employed successfully in order to control the
atomic motion \cite{ref12,ref13}. The cylindrical geometry is also important
for the investigation of fluorescence of substances in submicron capillaries
\cite{ref14,ref15}. A very important application of the decay rate theory in
the presence of dielectric fiber had been in the case of the photonic wire
lasers \cite{ref16}-\cite{ref19}. Finally, the circular cylindrical geometry
appears naturally at analysis of the carbon nanotubes \cite{ref20,ref21}.

The influence of a perfectly conducting cylindrical surface on decay rates
has been well investigated for an atom placed both inside a cylindrical
cavity \cite{ref22}-\cite{ref25} and near a perfectly conducting cylinder
\cite{ref26}. Spontaneous emission of an atom located inside a coaxial
waveguide was considered in \cite{ref27}.

The interaction of an atom with a dielectric, semiconductor, or a metallic
cylinder is a more complicated process. Van der Waals interaction between an
atom and a circular metallic nanowire was considered in \cite{ref28}. The
first investigation of the decay rate of an atom placed at the axis of a
circular dielectric fiber was undertaken within a classical approach in \cite%
{ref29,ref30}. Recently that problem attracted new interest \cite{ref31}-%
\cite{ref36}. In \cite{ref37,ref38} the spontaneous emission near carbon
nanotubes was considered.

Recently non-circular nanocylinders (the so called nanobelts or
nanoribbons) were successfully synthesized
\cite{ref39}-\cite{ref43}. Figure 1 illustrates the experimentally
obtained nanobelts \cite{ref39}.

The significance of studying such nanoobjects is very great because they
form the basis for the synthesis of all kinds of nanodevices, such as
nano-scale transducers, actuators, or sensors \cite{ref44}-\cite{ref47}. The
influence of such structures on optical characteristics of atoms and
molecules had not been yet considered, as far as we know.

The aim of this work is to investigate theoretically the influence of
dielectric and metallic cylinders of elliptical cross-section on the
spontaneous emission of atoms and molecules placed close to the cylinder.
This geometry is close to synthesized nanobelts and nanoribbons.

Though it is possible to separate variables in the Helmholtz equation within
the elliptic cylinder coordinates, still the electrodynamics and optics of
the dielectric elliptic cylinder remain insufficiently studied, due to
mathematical difficulties concerned with the Mathieu function \cite{ref48}.
In this paper we shall try to make up for this deficiency, and shall report
the investigation results dealing with the spontaneous emission of the atom
placed near an infinite elliptic cylinder.

Some elements of the theory for a spontaneous decay of an atom near material
bodies will be presented in section 2, where the main attention will be paid
to the nanobodies, whose dimensions are small as compared to the radiation
wavelength. In section 3, we report solution on a problem of spontaneous
decay of an atom near the perfectly conducting elliptic cylinder with
arbitrarily sized cross-section when retardation effects can be important.
In section 4, the radiative and nonradiative channels of a decaying atom
located near an elliptic nanocylinder, which is made of an arbitrary
material, will be considered analytically within the quasistatic approach.
The results obtained in sections 3 and 4 will be analyzed and illustrated
graphically in section 5. The geometry of the problem under study is
illustrated in Fig.2.

\section{Some elements of the theory for the spontaneous emission of an atom
placed near the nanobodies}

At weak coupling of an atom and a nanobody, i.e. in the case of the
exponential spontaneous decay, the expression for the linewidth $\gamma _{a}$
of excited level $a$ has the form \cite{ref49}:

\begin{equation}
\gamma _{a}=\gamma _{a,0}+\frac{{2}}{{\hbar }}\sum\limits_{n}\sum\limits_{%
\alpha ,\beta =1}^{3} d_{0,\alpha }^{an} d_{0,\beta}^{na}\text{Im}\left\{
G_{\alpha \beta }(\mathbf{r}^{\prime },\mathbf{r}^{\prime };\omega
_{na})\right\} \Theta \left( {\omega _{an}}\right)  \label{eq1}
\end{equation}

\noindent where $\gamma _{a,0}$ is the linewidth in free space without
nanobodies; $d_{0,\alpha }^{an}$, the matrix element of the dipole moment
operator between the states $a$ and $n$; $\omega _{an}=\left(
W_{a}-W_{n}\right) /\hbar $; and $G_{\alpha \beta }$, the reflected part of
the classical Green's function connected with the reflected field $E_{\alpha
}^{r}$ of a dipole $\mathbf{d}_{0}^{na}$ is placed at point $\mathbf{r}%
^{\prime }$ by the relation

\begin{equation}
E_{\alpha }^{r}\left( \mathbf{r}\right) =\sum\limits_{\beta
=1}^{3}d_{0,\beta }^{na}G_{\alpha \beta }(\mathbf{r},\mathbf{r}^{\prime
};\omega _{na})  \label{eq2}
\end{equation}

\noindent In Eqs. (\ref{eq1}) and (\ref{eq2}), $\alpha ,\beta =
1,2,3$ are the indices of the Cartesian axes, and $\Theta $ is the
Heaviside's unit-step function. Note that Eq.(\ref{eq1}) has been
obtained within the framework of the most general assumptions
(Fermi's golden rule and the fluctuation-dissipation theorem), and
has a very large domain of applicability. From Eq.(\ref{eq1}) it
is seen that total linewidth is the sum of partial widths, and
below we shall consider, for simplicity, the rate of transitions
between some two states only.

In the case of the nanobodies one can use the Rayleigh's perturbation long
wavelength theory. The Green's function of the reflected field may be
represented as the power series in the wave vector \cite{ref50}:

\begin{equation}
G_{\alpha \beta }\left( \mathbf{r}{,\mathbf{r}^{\prime };\omega }\right)
=G_{\alpha \beta }^{\left( {0}\right) }(\mathbf{r},\mathbf{r}^{\prime
})+kG_{\alpha \beta }^{\left( {1}\right) }(\mathbf{r},\mathbf{r}^{\prime
})+k^{2}G_{\alpha \beta }^{\left( {2}\right) }(\mathbf{r},\mathbf{r}^{\prime
})+ik^{3}G_{\alpha \beta }^{\left( {3}\right) }(\mathbf{r},\mathbf{r}%
^{\prime })+\ldots  \label{eq3}
\end{equation}

\noindent where $G_{\alpha \beta }^{\left( {j}\right) }$ ($j$=0, 1, 2,
\ldots ) are the coefficients that can be found from a solution of the
corresponding quasistatic problems; $k=\omega /c$. The first three terms (%
\ref{eq3}) describe the near fields, while the radiation fields emerge
starting from the $4^{\text{th}}$ term. By substituting Eq.(\ref{eq3}) into (%
\ref{eq1}) for the total rate of spontaneous decay $\gamma
_{e\longrightarrow g}$ from the state $e$ into $g$ near the nanobody, we
obtain the following expression:

\begin{eqnarray}
\gamma _{e\rightarrow g} &=&\underset{\text{nonradiative}}{\underbrace{\frac{%
2}{{\hbar }}\sum\limits_{\alpha ,\beta =1}^{3}d_{0,\alpha }^{eg}d_{0,\beta
}^{ge}\text{Im}\left\{ {G_{\alpha \beta }^{\left( {0}\right) }(\mathbf{r}%
^{\prime },\mathbf{r}^{\prime })+\ldots }\right\} }}  \notag \\
&&\underset{\text{radiative}}{\underbrace{\gamma _{0,e\rightarrow g}+{\frac{{%
2}}{{\hbar }}}{\sum\limits_{\alpha ,\beta =1}^{3}{d_{0,\alpha
}^{eg}d_{0,\beta }^{ge}\text{Re}{\left\{ {k^{3}G_{\alpha \beta }^{\left( {3}%
\right) }(\mathbf{r}^{\prime },\mathbf{r}^{\prime })+\ldots }\right\} }}}}}
\label{eq4}
\end{eqnarray}

\noindent Here $k\approx \omega _{0}/c$, and $\gamma _{0}$ and
$\omega _{0}$ are unperturbed by a nanobody the rate and frequency
of transitions between the states $a$ and $g$. The first term of
Eq.(\ref{eq4}) is nonzero for the absorbing media only, whereas
the rest of the terms are nonzero in the absence of absorption as
well (the dielectric or the perfectly conducting nanobodies).
Those terms describe mostly the radiative losses. Thus, to
determine the main terms of the nonradiative and radiative losses
it is
necessary to determine $G_{\alpha \beta }^{\left( {0}\right) }\left( \mathbf{%
r}^{\prime }{,}\mathbf{r}^{\prime }\right) $ and $G_{\alpha \beta }^{\left( {%
3}\right) }\left( \mathbf{r}^{\prime }{,}\mathbf{r}^{\prime }\right) $,
respectively.

It is a complex problem to determine directly the radiative losses described
by the 3-d order terms by $k$. But in the case of an atom localized close to
a nanobody the radiation is dipole-like, and the total dipole moment of the
atom + nanobody system could be found from $G_{\alpha \beta }^{\left( {0}%
\right) }\left( \mathbf{r}{,}\mathbf{r}^{\prime }\right) $ at large
distances from the system. Thus, in the case of the nanobodies the radiative
rate of the spontaneous transition (the transition from $e$ into $g)$ will
be described by \cite{ref35}:

\begin{equation}
\frac{\gamma ^{radiative}}{\gamma _{0}}=\frac{{{\left\vert \mathbf{d}{%
_{0}+\delta }\mathbf{d}\right\vert }^{2}}}{d_{0}^{2}},  \label{eq5}
\end{equation}

\noindent where $\mathbf{d}_{total}=\mathbf{d}_{0}+\delta \mathbf{d}$ is the
total dipole moment of the atom + nanobody system, and $\delta \mathbf{d}$
is the dipole moment induced in the nanobody. Here we will omit superscript
\textit{eg} in transition dipole moment $\mathbf{d}_{0}^{eg}$ and in $\gamma
_{e\longrightarrow g}$ and $\gamma _{0,e\longrightarrow g}$.

It should be noted that beside the radiative and nonradiative chanels of the
spontaneous decay for a dielectric cylinder (and for any dielectric
waveguided structure) one must take into account energy transfer into the
waveguided modes. In case of the dielectric (nonmetallic) nanobodies that
correction is exponentially small \cite{ref35} and may be neglected. But in
case of the complex dielectric permittivity of a nanocylinder (real metals),
the spontaneous decay waveguiding rate may be no longer a small value, and
that correction should be taken into account.

Thus, to describe the decay rate of an atom in the presence of a nanobody
one should determine the reflected field $\mathbf{E}^{r}$ in a quasistatic
approximation, and the dipole moment $\delta \mathbf{d}$ induced in the
nanobody. For a nanocylinder which size is comparable with radiation
wavelength one should use full set of Maxwell's equations to find the
reflected field.

\section{Spontaneous decay rate of an atom near a perfectly conducting
elliptic cylinder}

The problem of the spontaneous atomic decay near a perfectly conducting
elliptic cylinder of any geometry may be solved analytically in the elliptic
cylinder coordinates (Fig.3) ($1 \le u < \infty $, $- 1 \le v \le 1$, $-
\infty < z < \infty )$ \cite{ref48}

\begin{equation}
x=fuv,\quad y=f\sqrt{\left( u^{2}-1\right) \left( 1-v^{2}\right) },\quad z=z,
\label{eq6}
\end{equation}

\noindent where $f=\sqrt{a^{2}-b^{2}}$ is the half-distance between the
cylinder foci. Expression $u=u_{0}=a/f$ specifies the surface of the studied
cylinder.

Total field from the arbitrary current $\mathbf{j}\left( \mathbf{r}\right) $
and charge $\rho \left( \mathbf{r}\right) $ densities near the perfectly
conducting elliptic cylinder can be found from the equations \cite%
{ref51,ref52}

\begin{eqnarray}
E_{z}^{TM}\left( \mathbf{r}\right) &=&\int dr^{\prime }\left( ik\frac{%
j_{z}\left( \mathbf{r}^{\prime }\right) }{c}+\rho \left( \mathbf{r}^{\prime
}\right) \frac{{\partial }}{{\partial {z}^{\prime }}}\right) G^{TM}\left(
\mathbf{r}{,}\mathbf{r}^{\prime }\right) ,  \notag \\
H_{z}^{TE}\left( \mathbf{r}\right) &=&\frac{{1}}{{c}}\int dr^{\prime }\left[
\mathbf{j}{\left( \mathbf{r}^{\prime }\right) \times {\nabla }^{\prime }}%
\right] _{z}G^{TE}\left( \mathbf{r}{,}\mathbf{r}^{\prime }\right) ,
\label{eq7}
\end{eqnarray}

\noindent where the electric and magnetic Green's functions $G^{TM}$ and $%
G^{TE}$ are of the following form \cite{ref51} ($u>u^{\prime }>u_{0}$ )

\begin{eqnarray}
G^{TM}\left( \mathbf{r}{,}\mathbf{r}^{\prime }\right)
&=&2i\sum\limits_{\sigma =e,o}\sum\limits_{n=0}^{\infty }\int d\alpha
e^{i\alpha \left( {z-{z}^{\prime }}\right) }\frac{S\sigma _{n}\left( \beta
f,v\right) S\sigma _{n}\left( \beta f,v^{\prime }\right) }{M\sigma
_{n}\left( \beta f\right) }  \notag \\
&&\times H^{\left( {1}\right) }\sigma _{n}\left( \beta f,u\right) \left(
J\sigma _{n}\left( \beta f,u^{\prime }\right) -H^{\left( {1}\right) }\sigma
_{n}\left( \beta f,u^{\prime }\right) \right.  \notag \\
&&\times \left. \frac{J\sigma _{n}\left( \beta f,u_{0}\right) }{H^{\left(
1\right) }\sigma _{n}\left( \beta f,u_{0}\right) }\right) ,  \label{eq8}
\end{eqnarray}

\begin{eqnarray}
G^{TE}\left( \mathbf{r}{,}\mathbf{r}^{\prime }\right)
&=&2i\sum\limits_{\sigma =e,o}\sum\limits_{n=0}^{\infty }\int d\alpha
e^{i\alpha \left( {z-{z}^{\prime }}\right) }\frac{S\sigma _{n}\left( \beta
f,v\right) S\sigma _{n}\left( \beta f,v^{\prime }\right) }{M\sigma
_{n}\left( \beta f\right) }  \notag \\
&&\times H^{\left( {1}\right) }\sigma _{n}\left( \beta f,u\right) \left(
J\sigma _{n}\left( \beta f,u^{\prime }\right) -H^{\left( 1\right) }\sigma
_{n}\left( \beta f,u^{\prime }\right) \right.  \notag \\
&&\times \left. \left. \frac{\left( {\partial /\partial w}\right) J\sigma
_{n}\left( \beta f,w\right) }{{\left( {\partial /\partial w}\right) }%
H^{\left( {1}\right) }\sigma _{n}\left( \beta f,w\right) }\right\vert
_{w=u_{0}}\right) ,  \label{eq9}
\end{eqnarray}

\noindent where $\beta =\left( k^{2}-\alpha ^{2}\right) ^{1/2}$; $S\sigma
_{n}$ denote the \textquotedblleft angular\textquotedblright\ even ($\sigma
=e)$ and odd ($\sigma =o)$ Mathieu functions defined as follows \cite{ref48}:

\begin{eqnarray}
Se_{2n}\left( h,\cos \theta \right) &=&\sum\limits_{m=0}^{\infty
}Be_{2n}^{2m}\left( h\right) \cos \left( 2m\theta \right) ,  \notag \\
Se_{2n+1}\left( h,\cos \theta \right) &=&\sum\limits_{m=0}^{\infty
}Be_{2n+1}^{2m+1}\left( h\right) \cos \left( \left( 2m+1\right) \theta
\right) ,  \notag \\
So_{2n+1}\left( h,\cos \theta \right) &=&\sum\limits_{m=0}^{\infty
}Bo_{2n+1}^{2m+1}\left( h\right) \sin \left( \left( {2}m{+1}\right) \theta
\right) ,  \notag \\
So_{2n}\left( h,\cos \theta \right) &=&\sum\limits_{m=1}^{\infty
}Bo_{2n}^{2m}\left( h\right) \sin \left( 2m\theta \right) ,  \label{eq10}
\end{eqnarray}

\noindent where $B\sigma _{n}^{m}\left( h\right) $ are the expansion
coefficients ($Bo_{0}^{2m}\left( h\right) =0)$; $J\sigma _{n}$ and $%
H^{\left( {1}\right) }\sigma _{n}$ are the \textquotedblleft
radial\textquotedblright\ Mathieu functions, which are expressed through the
same coefficients $B\sigma _{n}^{m}\left( h\right) $. For example, functions
$J\sigma _{n}$ can be written down as \cite{ref48}:

\begin{eqnarray}
Je_{2n}\left( h,u\right) &=&\sqrt{\frac{\pi }{2}}\sum\limits_{m=0}^{\infty
}\left( {-1}\right) ^{n-m}Be_{2n}^{2m}\left( h\right) J_{2m}\left( hu\right)
,  \notag \\
Je_{2n+1}\left( h,u\right) &=&\sqrt{\frac{\pi }{{2}}}\sum\limits_{m=0}^{%
\infty }\left( {-1}\right) ^{n-m}Be_{2n+1}^{2m+1}\left( {h}\right)
J_{2m+1}\left( {hu}\right) ,  \notag \\
Jo_{2n+1}\left( h,u\right) &=&\sqrt{\frac{\pi }{2}}\frac{\sqrt{u^{2}-1}}{u}
\notag \\
&&\times \sum\limits_{m=0}^{\infty }\left( {-1}\right) ^{n-m}\left(
2m+1\right) Bo_{2n+1}^{2m+1}\left( h\right) J_{2m+1}\left( hu\right) ,
\notag \\
Jo_{2n}\left( h,u\right) &=&\sqrt{\frac{\pi }{2}}\frac{\sqrt{u^{2}-1}}{u}%
\sum\limits_{m=1}^{\infty }\left( {-1}\right) ^{n-m}2mBo_{2n}^{2m}\left( {h}%
\right) J_{2m}\left( hu\right) ,  \label{eq11}
\end{eqnarray}

\noindent where $J_{n}$ are the Bessel functions. For the functions $%
H^{\left( {1}\right) }\sigma _{n}$ the analogous expansions will be valid
provided that the Hankel functions are used (substitute $J_{m}\rightarrow
H_{m}^{\left( {1}\right) }$ in Eq.(\ref{eq11})). The normalization constant $%
M\sigma _{n}\left( h\right) $ is determined by the expression $M\sigma
_{n}\left( h\right) =\int\limits_{0}^{2\pi }d\theta \left[ S\sigma
_{n}\left( h,\cos \theta \right) \right] ^{2}$.

Integration over the longitudinal wave number $\alpha $ in Eqs. (\ref{eq8})
and (\ref{eq9}) is performed along the real axis assuming that the wave
number $k$ has a positive infinitely small imaginary addition ($k\rightarrow
k+i0^{+})$. It should also be noted that the first part of Eqs. (\ref{eq8})
and (\ref{eq9}) (which is proportional to $J\sigma _{n}\left( f\beta
,u^{\prime }\right) )$ corresponds to Green's function for free space (in
the absence of a cylinder), whereas the second part (which is proportional
to $H^{\left( {1}\right) }\sigma _{n}\left( f\beta ,u^{\prime }\right) )$
corresponds to the reflected field.

Note also that for a near-circular cylinder ($f\rightarrow 0$) one can
obtain from Eqs. (\ref{eq8}) and (\ref{eq9}) the respective expressions for
the Green's functions in cylindrical coordinates ($\rho ,\varphi $) if we
set that

\begin{eqnarray}
\frac{Se_{n}\left( f\beta ,v\right) }{\sqrt{Me_{n}\left( f\beta \right) }}%
Je_{n}\left( f\beta ,u\right) &\rightarrow &\frac{\cos \left( n\varphi
\right) }{\sqrt{2\left( 1+\delta _{0,n}\right) }}J_{n}\left( \beta \rho
\right) ,  \notag \\
\frac{So_{n}\left( f\beta ,v\right) }{\sqrt{Mo_{n}\left( f\beta \right) }}%
Jo_{n}\left( f\beta ,u\right) &\rightarrow &\frac{{\sin \left( n\varphi
\right) }}{\sqrt{2}}J_{n}\left( \beta \rho \right) ,  \label{eq12}
\end{eqnarray}

\noindent where $\delta _{0,n}$ is Kronecker's $\delta $-symbol.
Analogous expressions
for the functions $H^{\left( {1}\right) }\sigma _{n}$ are obtained from (\ref%
{eq12}) via a substitution $J_{n}\rightarrow H_{n}^{\left( {1}\right) }$.

The charge and current densities at excitation by a point dipole (atom) have
the form:

\begin{eqnarray}
\rho &=&-\left( \mathbf{d}{_{0}\nabla }\right) \delta \left( \mathbf{r}{-}%
\mathbf{r}^{\prime }\right) e^{-i\omega t},  \notag \\
j &=&-i\omega d_{0}\delta \left( \mathbf{r}{-}\mathbf{r}^{\prime }\right)
e^{-i\omega t},  \label{eq13}
\end{eqnarray}

\noindent where $\mathbf{d}_{0}$ is the transition dipole moment (below just
the dipole moment) of an atom placed at point $\mathbf{r}^{\prime }$ (below,
the primed coordinates will denote the atomic position). The exponential
term $e^{-i\omega t}$ will be omitted.

Knowing the field components (\ref{eq7}) we shall determine the rest of the
field components both for TM and TE modes \cite{ref53}

\begin{eqnarray}
E_{u,\alpha }^{TM} &=&\frac{i\alpha }{\beta ^{2}f}\sqrt{\frac{u^{2}-1}{%
u^{2}-v^{2}}}\frac{\partial E_{z,\alpha }^{TM}}{\partial u},\quad
E_{v,\alpha }^{TM}=\frac{i\alpha }{\beta ^{2}f}\sqrt{\frac{1-v^{2}}{%
u^{2}-v^{2}}}\frac{\partial E_{z,\alpha }^{TM}}{\partial v},  \notag \\
E_{u,\alpha }^{TE} &=&-\frac{ik}{\beta ^{2}f}\sqrt{\frac{1-v^{2}}{u^{2}-v^{2}%
}}\frac{\partial H_{z,\alpha }^{TE}}{\partial v},\quad E_{v,\alpha }^{TE}=%
\frac{ik}{\beta ^{2}f}\sqrt{\frac{u^{2}-1}{u^{2}-v^{2}}}\frac{\partial
H_{z,\alpha }^{TE}}{\partial u}.  \label{eq14}
\end{eqnarray}

\noindent In Eq.(\ref{eq14}) the $\alpha $ index denotes the
respective Fourier transform for the $z$ coordinate. Total field
is the sum of the TE and TM field components.

In the case of the dipole source (\ref{eq13}), the expressions (\ref{eq7})
for the fields may be rewritten in the following form:

\begin{eqnarray}
E_{z}^{TM}\left( \mathbf{r}\right) &=&k^{2}d_{0,z}G^{TM}\left( \mathbf{r,r}%
^{\prime }\right) +\left( \mathbf{d}{_{0}{\nabla }^{\prime }}\right) \frac{{%
\partial }}{{\partial {z}^{\prime }}}G^{TM}\left( \mathbf{r,r}^{\prime
}\right) ,  \notag \\
H_{z}^{TE}\left( \mathbf{r}\right) &=&-ik\left[ \mathbf{d}{_{0}}\times {{%
\nabla }^{\prime }}\right] _{z}G^{TE}\left( \mathbf{r,r}^{\prime }\right) .
\label{eq15}
\end{eqnarray}

By using (\ref{eq8}), (\ref{eq9}), and (\ref{eq14}), (\ref{eq15}) one can
represent the spontaneous emission rate (\ref{eq1}) as the sum of the TM and
TE contributions:

\begin{equation}
\frac{\gamma }{\gamma _{0}}=1-\frac{{3}}{{2}}\text{Re}\left(
W^{TM}+W^{TE}\right) .  \label{eq16}
\end{equation}

\noindent In some particular cases Eq.(\ref{eq16}) will take the
next relatively simple form.

\begin{center}
\textit{Dipole moment is oriented along x-axis (}$\mathbf{d}_{0}=d_{0}%
\mathbf{e}_{x}$\textit{)}
\end{center}

A) Atom is placed on the $x$-axis ($\mathbf{r}^{\prime }={x}^{\prime }%
\mathbf{e}_{x}$ )

\begin{eqnarray}
W_{x}^{TM} &=&4\sum\limits_{n=0}^{\infty }\int\limits_{0}^{k}d\alpha \frac{%
\alpha ^{2}}{k^{3}\left( \beta f\right) ^{2}}\frac{{1}}{Me_{n}\left( \beta
f\right) }  \notag \\
&&\times \left[ \frac{{\partial }}{\partial u^{\prime }}H^{\left( {1}\right)
}e_{n}\left( \beta f,u^{\prime }\right) \right] ^{2}\frac{Je_{n}\left( \beta
f,u_{0}\right) }{H^{\left( {1}\right) }e_{n}\left( \beta f,u_{0}\right) },
\label{eq17}
\end{eqnarray}

\begin{eqnarray}
W_{x}^{TE} &=&\frac{{4}}{{{u}^{\prime }{}^{2}-1}}\sum\limits_{n=0}^{\infty
}\int\limits_{0}^{k}\frac{d\alpha }{k\left( \beta f\right) ^{2}}\frac{{1}}{%
Mo_{n}\left( \beta f\right) }  \notag \\
&&\times \left[ H^{\left( {1}\right) }o_{n}\left( \beta f,u^{\prime }\right) %
\right] ^{2}\left. \frac{\left( \partial /\partial w\right) Jo_{n}\left(
\beta f,w\right) }{\left( {\partial /}\partial w\right) H^{\left( {1}\right)
}o_{n}\left( \beta f,w\right) }\right\vert _{w=u_{0}}.  \label{eq18}
\end{eqnarray}

B) Atom is placed on the $y$-axis ($\mathbf{r}^{\prime }={y}^{\prime }%
\mathbf{e}_{y}$ )

\begin{eqnarray}
W_{x}^{TM} &=&\frac{{4}}{u^{\prime }{}^{2}}\sum\limits_{n=0}^{\infty
}\int\limits_{0}^{k}d\alpha \frac{\alpha ^{2}}{k^{3}\left( \beta f\right)
^{2}}\frac{\left[ \left. \left( \partial /\partial v^{\prime }\right)
Se_{2n+1}\left( \beta f,v^{\prime }\right) \right\vert _{v^{\prime }=0}%
\right] ^{2}}{Me_{2n+1}\left( \beta f\right) }  \notag \\
&&\times \left[ H^{\left( {1}\right) }e_{2n+1}\left( \beta f,u^{\prime
}\right) \right] ^{2}\frac{Je_{2n+1}\left( \beta f,u_{0}\right) }{H^{\left( {%
1}\right) }e_{2n+1}\left( \beta f,u_{0}\right) }  \notag \\
&&+\frac{{4}}{u^{\prime }{}^{2}}\sum\limits_{n=0}^{\infty
}\int\limits_{0}^{k}d\alpha \frac{\alpha ^{2}}{k^{3}\left( \beta f\right)
^{2}}\frac{\left[ \left. \left( \partial /\partial v^{\prime }\right)
So_{2n}\left( \beta f,v^{\prime }\right) \right\vert _{v^{\prime }=0}\right]
^{2}}{Mo_{2n}\left( \beta f\right) }  \notag \\
&&\times \left[ H^{\left( {1}\right) }o_{2n}\left( \beta f,u^{\prime
}\right) \right] ^{2}\frac{Jo_{2n}\left( \beta f,u_{0}\right) }{H^{\left( {1}%
\right) }o_{2n}\left( \beta f,u_{0}\right) },  \label{eq19}
\end{eqnarray}

\begin{eqnarray}
W_{x}^{TE} &=&4\left( {1-{\frac{{1}}{{{u}^{\prime }{}^{2}}}}}\right)
\sum\limits_{n=0}^{\infty }\int\limits_{0}^{k}\frac{d\alpha }{k\left( \beta
f\right) ^{2}}\frac{\left[ Se_{2n}\left( \beta f,0\right) \right] ^{2}}{%
Me_{2n}\left( \beta f\right) }  \notag \\
&&\times \left[ \frac{{\partial }}{\partial u^{\prime }}H^{\left( {1}\right)
}e_{2n}\left( \beta f,u^{\prime }\right) \right] ^{2}\left. \frac{\left(
\partial /\partial w\right) Je_{2n}\left( \beta f,w\right) }{\left( \partial
/\partial w\right) H^{\left( {1}\right) }e_{2n}\left( \beta f,w\right) }%
\right\vert _{w=u_{0}}  \notag \\
&&+4\left( 1-\frac{{1}}{u^{\prime }{}^{2}}\right) \sum\limits_{n=0}^{\infty
}\int\limits_{0}^{k}\frac{d\alpha }{k\left( \beta f\right) ^{2}}\frac{\left[
So_{2n+1}\left( \beta f,0\right) \right] {^{2}}}{Mo_{2n+1}\left( \beta
f\right) }  \notag \\
&&\times \left[ \frac{{\partial }}{\partial u^{\prime }}H^{\left( {1}\right)
}o_{2n+1}\left( \beta f,u^{\prime }\right) \right] ^{2}\left. \frac{\left( {%
\partial /}\partial w\right) Jo_{2n+1}\left( \beta f,w\right) }{\left( {%
\partial /}\partial w\right) H^{\left( {1}\right) }o_{2n+1}\left( \beta
f,w\right) }\right\vert _{w=u_{0}}.  \label{eq20}
\end{eqnarray}

\begin{center}
\textit{\ Dipole moment is oriented along y-axis (}$\mathbf{d}_{0}=d_{0}%
\mathbf{e}_{y}$\textit{)}
\end{center}

A) Atom is placed on the x-axis ($\mathbf{r}^{\prime }=x^{\prime }\mathbf{e}%
_{x}$ )

\begin{eqnarray}
W_{y}^{TM} &=&\frac{{4}}{u^{\prime }{}^{2}-1}\sum\limits_{n=0}^{\infty
}\int\limits_{0}^{k}d\alpha \frac{\alpha ^{2}}{k^{3}\left( \beta f\right)
^{2}}\frac{{1}}{Mo_{n}\left( \beta f\right) }  \notag \\
&&\times \left[ H^{\left( {1}\right) }o_{n}\left( \beta f,u^{\prime }\right) %
\right] ^{2}\frac{Jo_{n}\left( \beta f,u_{0}\right) }{H^{\left( {1}\right)
}o_{n}\left( \beta f,u_{0}\right) },  \label{eq21}
\end{eqnarray}

\begin{eqnarray}
W_{y}^{TE} &=&4\sum\limits_{n=0}^{\infty }\int\limits_{0}^{k}\frac{d\alpha }{%
k\left( \beta f\right) ^{2}}\frac{{1}}{Me_{n}\left( \beta f\right) }  \notag
\\
&&\times \left[ \frac{{\partial }}{\partial u^{\prime }}H^{\left( {1}\right)
}e_{n}\left( \beta f,u^{\prime }\right) \right] ^{2}  \notag \\
&&\times \left. \frac{\left( \partial /\partial w\right) Je_{n}\left( \beta
f,w\right) }{\left( \partial /\partial w\right) H^{\left( {1}\right)
}e_{n}\left( \beta f,w\right) }\right\vert _{w=u_{0}}.  \label{eq22}
\end{eqnarray}

B) Atom is placed on the y-axis ($\mathbf{r}^{\prime }=y^{\prime }\mathbf{e}%
_{y}$ )

\begin{eqnarray}
W_{y}^{TM} &=&4\left( 1-\frac{{1}}{u^{\prime }{}^{2}}\right)
\sum\limits_{n=0}^{\infty }\int\limits_{0}^{k}d\alpha \frac{\alpha ^{2}}{%
k^{3}\left( \beta f\right) ^{2}}\frac{\left[ Se_{2n}\left( \beta f,0\right) %
\right] {^{2}}}{Me_{2n}\left( \beta f\right) }  \notag \\
&&\times \left[ \frac{{\partial }}{\partial u^{\prime }}H^{\left( {1}\right)
}e_{2n}\left( \beta f,u^{\prime }\right) \right] ^{2}\frac{Je_{2n}\left(
\beta f,u_{0}\right) }{H^{\left( {1}\right) }e_{2n}\left( \beta
f,u_{0}\right) }  \notag \\
&&+4\left( 1-\frac{{1}}{u^{\prime }{}^{2}}\right) \sum\limits_{n=0}^{\infty
}\int\limits_{0}^{k}d\alpha \frac{\alpha ^{2}}{k^{3}\left( \beta f\right)
^{2}}\frac{\left[ So_{2n+1}\left( \beta f,0\right) \right] {^{2}}}{%
Mo_{2n+1}\left( \beta f\right) }  \notag \\
&&\times \left[ \frac{{\partial }}{\partial u^{\prime }}H^{\left( {1}\right)
}o_{2n+1}\left( \beta f,u^{\prime }\right) \right] ^{2}\frac{Jo_{2n+1}\left(
\beta f,u_{0}\right) }{H^{\left( {1}\right) }o_{2n+1}\left( \beta
f,u_{0}\right) },  \label{eq23}
\end{eqnarray}

\begin{eqnarray}
W_{y}^{TE} &=&\frac{{4}}{u^{\prime }{}^{2}}\sum\limits_{n=0}^{\infty
}\int\limits_{0}^{k}\frac{d\alpha }{k\left( \beta f\right) ^{2}}\frac{\left[
\left. \left( \partial /\partial v^{\prime }\right) Se_{2n+1}\left( \beta
f,v^{\prime }\right) \right\vert _{v^{\prime }=0}\right] ^{2}}{%
Me_{2n+1}\left( \beta f\right) }  \notag \\
&&\times \left[ H^{\left( {1}\right) }e_{2n+1}\left( \beta f,u^{\prime
}\right) \right] ^{2}\left. \frac{\left( {\partial }/\partial w\right)
Je_{2n+1}\left( \beta f,w\right) }{\left( {\partial }/\partial w\right)
H^{\left( {1}\right) }e_{2n+1}\left( \beta f,w\right) }\right\vert _{w=u_{0}}
\notag \\
&&+\frac{{4}}{u^{\prime }{}^{2}}\sum\limits_{n=0}^{\infty
}\int\limits_{0}^{k}\frac{d\alpha }{k\left( \beta f\right) ^{2}}\frac{\left[
\left. \left( \partial /\partial v^{\prime }\right) So_{2n}\left( \beta
f,v^{\prime }\right) \right\vert _{v^{\prime }=0}\right] ^{2}}{Mo_{2n}\left(
\beta f\right) }  \notag \\
&&\times \left[ H^{\left( {1}\right) }o_{2n}\left( \beta f,u^{\prime
}\right) \right] ^{2}\left. \frac{\left( \partial /\partial w\right)
Jo_{2n}\left( \beta f,w\right) }{{\left( \partial /\partial w\right) }%
H^{\left( {1}\right) }o_{2n}\left( \beta f,w\right) }\right\vert _{w=u_{0}}.
\label{eq24}
\end{eqnarray}

\begin{center}
\textit{Dipole moment is oriented along z-axis (}$\mathbf{d}_{0}=d_{0}%
\mathbf{e}_{z}$\textit{) }
\end{center}

A) Atom is placed on the $x$-axis ($\mathbf{r}^{\prime }=x^{\prime }\mathbf{e%
}_{x}$ )

\begin{eqnarray}
W_{z}^{TM} &=&4\sum\limits_{n=0}^{\infty }\int\limits_{0}^{k}d\alpha \frac{%
\beta ^{2}}{k^{3}}\frac{{1}}{Me_{n}\left( \beta f\right) }\left[ H^{\left( {1%
}\right) }e_{n}\left( \beta f,u^{\prime }\right) \right] ^{2}\frac{%
Je_{n}\left( \beta f,u_{0}\right) }{H^{\left( {1}\right) }e_{n}\left( \beta
f,u_{0}\right) },  \notag \\
W_{z}^{TE} &=&0.  \label{eq25}
\end{eqnarray}

B) Atom is placed on the $y$-axis ($\mathbf{r}^{\prime }=y^{\prime }\mathbf{e%
}_{y}$ )

\begin{eqnarray}
W_{z}^{TM} &=&4\sum\limits_{n=0}^{\infty }\int\limits_{0}^{k}d\alpha \frac{%
\beta ^{2}}{k^{3}}\frac{\left[ Se_{2n}\left( \beta f,0\right) \right] {^{2}}%
}{Me_{2n}\left( \beta f\right) }\left[ H^{\left( {1}\right) }e_{2n}\left(
\beta f,u^{\prime }\right) \right] ^{2}\frac{Je_{2n}\left( \beta
f,u_{0}\right) }{H^{\left( {1}\right) }e_{2n}\left( \beta f,u_{0}\right) }
\notag \\
&&+4\sum\limits_{n=0}^{\infty }\int\limits_{0}^{k}d\alpha \frac{\beta ^{2}}{%
k^{3}}\frac{\left[ So_{2n+1}\left( \beta f,0\right) \right] {^{2}}}{%
Mo_{2n+1}\left( \beta f\right) }\left[ H^{\left( {1}\right) }o_{2n+1}\left(
\beta f,u^{\prime }\right) \right] ^{2}  \notag \\
&&\times \frac{Jo_{2n+1}\left( \beta f,u_{0}\right) }{H^{\left( {1}\right)
}o_{2n+1}\left( \beta f,u_{0}\right) },  \notag \\
W_{z}^{TE} &=&0.  \label{eq26}
\end{eqnarray}

\begin{center}
\textit{Case of low ellipticity (}$f\rightarrow 0$\textit{) }
\end{center}

If a cylinder has low ellipticity ($f\rightarrow 0)$ one can find explicit
expressions for the spontaneous emission rate (\ref{eq16}) in the form of
the power series $f$. That can be done if the expansion for the coefficients
$B\sigma _{n}^{m}\left( h\right) $ at $h\rightarrow 0$ is known. Keeping the
terms of the second-order smallness by $h$ on can write down the expressions
for $B\sigma _{n}^{m}\left( h{\rightarrow 0}\right) $in the following form ($%
n=1,2,3,\ldots $):

\begin{eqnarray}
Be_{0}^{0} &\approx &1+\frac{h^{2}}{{8}},\quad Be_{0}^{2}\approx -\frac{h^{2}%
}{{8}},  \notag \\
Be_{2n}^{2\left( n-1\right) } &\approx &\frac{h^{2}}{16\left( 2n-1\right) }%
,\quad Be_{2n}^{2n}\approx 1-\frac{h^{2}}{8\left( {4}n^{2}{-1}\right) },
\notag \\
Be_{2n}^{2\left( n+1\right) } &\approx &-\frac{h^{2}}{16\left( 2n+1\right) },
\label{eq27}
\end{eqnarray}

\begin{eqnarray}
Be_{1}^{1} &\approx &1+\frac{h^{2}}{{32}},\quad Be_{1}^{3}\approx -\frac{%
h^{2}}{{32}},  \notag \\
Be_{2n+1}^{2n-1} &\approx &\frac{h^{2}}{32n},\quad Be_{2n+1}^{2n+1}\approx 1-%
\frac{h^{2}}{32n\left( n+1\right) },  \notag \\
Be_{2n+1}^{2n+3} &\approx &-\frac{h^{2}}{32\left( n+1\right) },  \label{eq28}
\end{eqnarray}

\noindent and

\begin{eqnarray}
Bo_{1}^{1} &\approx &1+\frac{3h^{2}}{{32}},\quad Bo_{1}^{3}\approx -\frac{%
h^{2}}{{32}},  \notag \\
Bo_{2n+1}^{2n-1} &\approx &\frac{h^{2}}{32n\left( 2n{+1}\right) },\quad
Bo_{2n+1}^{2n+1}\approx \frac{{1}}{2n+1}\left( 1+\frac{h^{2}}{32n\left(
n+1\right) }\right) ,  \notag \\
Bo_{2n+1}^{2n+3} &\approx &-\frac{h^{2}}{32\left( n+1\right) \left(
2n+1\right) },  \label{eq29}
\end{eqnarray}

\begin{eqnarray}
Bo_{2n}^{2\left( n-1\right) } &\approx &\frac{h^{2}}{32n\left( 2n-1\right) }%
,\quad Bo_{2n}^{2n}\approx \frac{{1}}{2n}\left( 1+\frac{h^{2}}{{8}\left(
4n^{2}-1\right) }\right) ,  \notag \\
Bo_{2n}^{2\left( n+1\right) } &\approx &-\frac{h^{2}}{32n\left( 2n{+1}%
\right) }.  \label{eq30}
\end{eqnarray}

\noindent The rest of the coefficients $Be_{n}^{m}$ are either
zero or have a higher
infinitesimal order. By substituting Eqs. (\ref{eq27})-(\ref{eq30}) into (%
\ref{eq10}) and (\ref{eq11}), and assuming that in the cylindrical
coordinates $\left( \rho ,\varphi \right) $ we have $fu\approx \rho \left( {%
1+{\frac{{1}}{{2}}}\left( {{\frac{{f}}{{\rho }}}}\right) ^{2}\sin
^{2}\varphi }\right) $ and $v\approx \cos \varphi \left( 1-\frac{{1}}{{2}}%
\left( {{\frac{{f}}{{\rho }}}}\right) ^{2}\sin ^{2}\varphi \right) $, we can
find expansions in the region of $\ f\rightarrow 0$ for the Mathieu
functions and their derivatives with accuracy up to $f^{2}$. By restricting
ourselves to the main expansion terms only, and assuming that $fu^{\prime
}=\rho ^{\prime }$ and $fu_{0}=a$ \ we obtain the following expressions for $%
W^{TM}$ and $W^{TE}$.

A) Dipole moment is oriented along $\rho $-coordinate line

\begin{eqnarray}
W_{\rho }^{TM} &\approx &\sum\limits_{n=0}^{\infty }\left( {2-}\delta
_{0,n}\right) \int\limits_{0}^{k}d\alpha \frac{\alpha ^{2}}{k^{3}}\left[
\left. \frac{d}{dz}H_{n}^{\left( {1}\right) }\left( z\right) \right\vert
_{z=\beta {\rho }^{\prime }}\right] ^{2}\frac{J_{n}\left( \beta a\right) }{%
H_{n}^{\left( {1}\right) }\left( \beta a\right) }+O\left( f^{2}\right) ,
\notag \\
W_{\rho }^{TE} &\approx &2\sum\limits_{n=1}^{\infty }n^{2}\int\limits_{0}^{k}%
\frac{d\alpha }{k\left( \beta \rho ^{\prime }\right) ^{2}}\left[
H_{n}^{\left( {1}\right) }\left( \beta \rho ^{\prime }\right) \right]
^{2}\left. \frac{\left( d/dz\right) J_{n}\left( z\right) }{\left(
d/dz\right) H_{n}^{\left( {1}\right) }\left( z\right) }\right\vert _{z=\beta
a}  \notag \\
&&+O\left( f^{2}\right) .  \label{eq31}
\end{eqnarray}

B) Dipole moment is oriented along $\varphi $-coordinate line

\begin{eqnarray}
W_{\varphi }^{TM} &\approx &2\sum\limits_{n=1}^{\infty
}n^{2}\int\limits_{0}^{k}d\alpha \frac{\alpha ^{2}}{k^{3}\left( {\beta {\rho
}^{\prime }}\right) ^{2}}\left[ H_{n}^{\left( {1}\right) }\left( \beta \rho
^{\prime }\right) \right] ^{2}\frac{J_{n}\left( \beta a\right) }{%
H_{n}^{\left( {1}\right) }\left( \beta a\right) }+O\left( f^{2}\right) ,
\notag \\
W_{\varphi }^{TE} &\approx &\sum\limits_{n=0}^{\infty }\left( 2-\delta
_{0,n}\right) \int\limits_{0}^{k}\frac{d\alpha }{k}\left[ \left. \frac{d}{dz}%
H_{n}^{\left( {1}\right) }\left( z\right) \right\vert _{z=\beta {\rho }%
^{\prime }}\right] ^{2}  \notag \\
&&\times \left. \frac{\left( d/dz\right) J_{n}\left( z\right) }{\left(
d/dz\right) H_{n}^{\left( {1}\right) }\left( z\right) }\right\vert _{z=\beta
a}+O\left( f^{2}\right) .  \label{eq32}
\end{eqnarray}

C) Dipole moment is oriented along $z$-axis

\begin{eqnarray}
W_{z}^{TM} &\approx &\sum\limits_{n=0}^{\infty }\left( 2-\delta
_{0,n}\right) \int\limits_{0}^{k}d\alpha \frac{\beta ^{2}}{k^{3}}\left[
H_{n}^{\left( {1}\right) }\left( \beta \rho ^{\prime }\right) \right] ^{2}%
\frac{J_{n}\left( \beta a\right) }{H_{n}^{\left( {1}\right) }\left( \beta
a\right) }+O\left( f^{2}\right) ,  \notag \\
W_{z}^{TE} &=&0.  \label{eq33}
\end{eqnarray}

Note that the main terms (\ref{eq31})-(\ref{eq33}) coincide as in the case
of the analogous expressions for $W_{\rho} $, $W_{\varphi} $, and $W_{z} $
in the circular cylinder of $a$ radius \cite{ref26}. This circumstance
supports the validity of our calculations.

\begin{center}
\textit{Atom is placed in close proximity to surface of a cylinder}
\end{center}

In the most interesting case of an atom placed in close proximity to the
surface of a cylinder ($u^{\prime }\rightarrow u_{0}$ ). The decay rate of a
tangentially to surface oriented dipole tends to zero.\ The nonzero rate of
decays is possible if the dipole moment orientation is normal to the
cylindrical surface, i.e. for the dipole that is oriented along the $u$%
-coordinate line

\begin{eqnarray}
W_{u}^{TM} &=&4\left( \frac{u_{0}^{2}-1}{u_{0}^{2}-v^{\prime }{}^{2}}\right)
\sum\limits_{\sigma =e,o}\sum\limits_{n=0}^{\infty
}\int\limits_{0}^{k}d\alpha \frac{\alpha ^{2}}{k^{3}\left( \beta f\right)
^{2}}\frac{\left[ S\sigma _{n}\left( \beta f,v^{\prime }\right) \right] {^{2}%
}}{M\sigma _{n}\left( {\beta f}\right) }  \notag \\
&&\times \left[ \left. \frac{{\partial }}{\partial u^{\prime }}H^{\left( {1}%
\right) }\sigma _{n}\left( \beta f,u^{\prime }\right) \right\vert
_{u^{\prime }=u_{0}}\right] ^{2}\frac{J\sigma _{n}\left( \beta
f,u_{0}\right) }{H^{\left( {1}\right) }\sigma _{n}\left( \beta
f,u_{0}\right) },  \notag \\
W_{u}^{TE} &=&4\left( \frac{1-v^{\prime }{}^{2}}{u_{0}^{2}-v^{\prime }{}^{2}}%
\right) \sum\limits_{\sigma =e,o}\sum\limits_{n=0}^{\infty
}\int\limits_{0}^{k}\frac{d\alpha }{k\left( \beta f\right) ^{2}}\frac{\left[
\left( \partial /\partial v^{\prime }\right) S\sigma _{n}\left( \beta
f,v^{\prime }\right) \right] {^{2}}}{M\sigma _{n}\left( \beta f\right) }
\notag \\
&&\times \left[ H^{\left( {1}\right) }\sigma _{n}\left( \beta f,u_{0}\right) %
\right] ^{2}\left. \frac{\left( {\partial /}\partial w\right) J\sigma
_{n}\left( \beta f,w\right) }{\left( {\partial /}\partial w\right) H^{\left(
{1}\right) }\sigma _{n}\left( \beta f,w\right) }\right\vert _{w=u_{0}}.
\label{eq34}
\end{eqnarray}

When elliptic cylinder radius is tending to zero $kfu_{0}\rightarrow 0$, the
decay rate (\ref{eq34}) of a tangentially to surface oriented dipole tends
to zero the main contribution to the decay rates of normally oriented dipole
is due to the terms with $n=0$ and $n=1$ of (\ref{eq34}). By expanding the
expressions (\ref{eq10}) and (\ref{eq11}) into series at $%
kf,\;kfu_{0}\rightarrow 0$ (the procedure of finding the expansions for the
coefficients $B\sigma _{0}^{m}$ and $B\sigma _{1}^{m}$ is described in
detail in \cite{ref48}) one can obtain from (\ref{eq16}) and (\ref{eq34})
the next simple asymptotic formula:

\begin{eqnarray}
\left. \left( \frac{\gamma }{\gamma _{0}}\right) _{u}\right\vert
_{kfu^{\prime }=kfu_{0}\rightarrow 0} &=&\frac{{3}}{2\left(
u_{0}^{2}-v^{\prime }{}^{2}\right) \left( kf\right) ^{2}}  \notag \\
&&\times \left( 1+\frac{{2}}{{\pi }}\arctan L^{\ast }\left( u_{0}\right) +%
\frac{{4\left( \ln 2-1\right) }}{{\pi ^{2}}\left( 1+L^{\ast }{}^{2}\left(
u_{0}\right) \right) }+\ldots \right)  \notag \\
&&+\frac{\left( u_{0}+\sqrt{u_{0}^{2}-1}\right) ^{2}}{u_{0}^{2}-v^{\prime
}{}^{2}}+\ldots  \label{eq35}
\end{eqnarray}

\noindent in which

\begin{equation}
L^{\ast }\left( u_{0}\right) =\frac{{2}}{\pi }\left\{ {\ln }\left[ \frac{kf}{%
{4}}\left( u_{0}+\sqrt{u_{0}^{2}-1}\right) \right] {+}\gamma _{E}\right\} ,
\label{eq36}
\end{equation}

\noindent where $\gamma _{E}\approx 0.57722$ is the Euler's constant. In a
particular case of a circular cylinder of $a$-radius ($f\rightarrow 0$, $\
fu_{0}=a)$ we obtain from (\ref{eq35}) the expansion that coincides with
analogous asymptotic expression for the circular cylinder \cite{ref26}

\begin{equation}
\left. \left( \frac{\gamma }{\gamma _{0}}\right) _{\rho }\right\vert _{k\rho
^{\prime }=ka\rightarrow 0}=\frac{{3}}{{2}\left( ka\right) ^{2}}\left( {1+{%
\frac{{2}}{\pi }}\arctan }L^{\ast }{+}\frac{{4\left( {\ln 2-1}\right) }}{\pi
^{2}\left( {1+}L^{\ast }{}^{2}\right) }{+\ldots }\right) +4+\ldots
\label{eq37}
\end{equation}

\noindent where $L^{\ast }=\frac{{2}}{\pi }\left[ \ln \left( \frac{ka}{{2}}%
\right) {+}\gamma _{E}\right] $.

In the most interesting case of a very thin nanobelt ($u_{0}\rightarrow 1$)
the decay rate of the atom (having normal orientation of the dipole moment)
can tend both to infinity (the atom is at the nanobelt edge ($v^{\prime
}=\pm 1$)) and take finite values at other positions of the atom. This
signifies nontrivial character of the found expressions (\ref{eq35}).

\section{ Spontaneous decay rate of an atom near elliptic nanocylinder made
of arbitrary material}

In the case of the dielectric or metallic cylinder of arbitrary dimensions
and arbitrary permittivity $\varepsilon $ the problem of the spontaneous
decay rate of an atom may also be solved analytically \cite{ref54}. But such
a solution is so cumbersome that it is very difficult to use it practically.
In this section we restrict ourselves to a practically important case of a
nanofiber or a metal nanowire of elliptical cross-section. It may be easy to
describe the spontaneous decay, in this case.

\subsection{ The rate of the radiative decays}

In order to find the spontaneous radiative decay rate of an atom placed at
point $\mathbf{r}^{\prime }$ near to elliptic nanocylinder of arbitrary
material it is necessary to calculate a dipole moment induced in a
nanocylinder, in accordance with Eq.(\ref{eq5}). In order to find the dipole
moment one may use the found analytical solution \cite{ref11} for three-axis
ellipsoid and turn one of the semi-axes to the infinity (without using the
elliptic cylinder coordinates). As a result, the expression for the induced
dipole moment on the elliptic cylinder made of material with the dielectric
constant $\varepsilon $ with semi-axes $a$ and $b$ on the $x$- and $y$-axes,
respectively (Fig.2), will take the form:

\begin{eqnarray}
\delta \mathbf{d} &=&-\frac{{3}}{{2}}\left( d_{0,x}P_{xx}F_{a}\left( \xi
^{\prime }\right) \mathbf{e}_{x}+d_{0,y}P_{yy}F_{b}\left( \xi ^{\prime
}\right) \mathbf{e}_{y}\right)  \notag \\
&&+\frac{{3}}{\sqrt{\left( a^{2}+\xi ^{\prime }\right) \left( b^{2}+\xi
^{\prime }\right) }}\left( \frac{d_{0,x}x^{\prime }}{a^{2}+\xi ^{\prime }}+%
\frac{d_{0,y}y^{\prime }}{b^{2}+\xi ^{\prime }}\right)  \notag \\
&&\times \left( \frac{P_{xx}x^{\prime }}{a^{2}+\xi ^{\prime }}\mathbf{e}_{x}+%
\frac{P_{xx}y^{\prime }}{b^{2}+\xi ^{\prime }}\mathbf{e}_{y}\right) \left(
\frac{x^{\prime }{}^{2}}{\left( a^{2}+\xi ^{\prime }\right) ^{2}}+\frac{%
y^{\prime }{}^{2}}{\left( b^{2}+\xi ^{\prime }\right) ^{2}}\right) ^{-1},
\label{eq38}
\end{eqnarray}

\noindent where the components of the polarizability tensor of an elliptic
nanocylinder are:

\begin{eqnarray}
P_{xx} &=&{\frac{{1}}{{3}}}ab\left( a+b\right) \frac{\varepsilon -1}{%
a+\varepsilon b},  \notag \\
P_{yy} &=&\frac{{1}}{{3}}ab\left( a+b\right) \frac{\varepsilon -1}{%
\varepsilon a+b},  \label{eq39}
\end{eqnarray}

\noindent where $\xi ^{\prime }$ is the positive root of equation $\frac{%
x^{\prime }{}^{2}}{a^{2}+\xi ^{\prime }}+\frac{y^{\prime }{}^{2}}{b^{2}+\xi
^{\prime }}=1$;

\begin{eqnarray}
F_{a}\left( \xi ^{\prime }\right) &=&\frac{{2}}{\sqrt{a^{2}+\xi ^{\prime }}%
\left( \sqrt{a^{2}+\xi ^{\prime }}+\sqrt{b^{2}+\xi ^{\prime }}\right) },
\notag \\
F_{b}\left( \xi ^{\prime }\right) &=&\frac{{2}}{\sqrt{b^{2}+\xi ^{\prime }}%
\left( \sqrt{a^{2}+\xi ^{\prime }}+\sqrt{b^{2}+\xi ^{\prime }}\right) }.
\label{eq40}
\end{eqnarray}

One can obtain the following expressions for some particular cases of the
dipole moment orientation along the Cartesian axes.

A) Dipole moment is oriented along $x$-axis ($\mathbf{d}_{0}=d_{0}\mathbf{e}%
_{x}$ )

\begin{eqnarray}
\delta \mathbf{d} &=&-\frac{d_{0}\left( \varepsilon -1\right) ab\left(
a+b\right) \mathbf{e}_{x}}{\sqrt{a^{2}+{\xi }^{\prime }}\left( {\sqrt{a^{2}+{%
\xi }^{\prime }}+\sqrt{b^{2}+{\xi }^{\prime }}}\right) \left( {a+\varepsilon
b}\right) }  \notag \\
&&+\frac{d_{0}\left( {\varepsilon -1}\right) ab\left( a+b\right) x^{\prime }%
}{\left( a^{2}+\xi ^{\prime }\right) ^{3/2}\sqrt{b^{2}+\xi ^{\prime }}}%
\left( \frac{x^{\prime }{}^{2}}{\left( a^{2}+\xi ^{\prime }\right) ^{2}}+%
\frac{y^{\prime }{}^{2}}{\left( b^{2}+\xi ^{\prime }\right) ^{2}}\right)
^{-1}  \notag \\
&&\times \left\{ \frac{x^{\prime }\mathbf{e}_{x}}{\left( a^{2}+\xi ^{\prime
}\right) \left( a+\varepsilon b\right) }+\frac{y^{\prime }\mathbf{e}_{y}}{%
\left( b^{2}+\xi ^{\prime }\right) \left( \varepsilon a+b\right) }\right\} .
\label{eq41}
\end{eqnarray}

B) Dipole moment is oriented along $y$-axis ($\mathbf{d}_{0}=d_{0}\mathbf{e}%
_{y}$ )

\begin{eqnarray}
\delta \mathbf{d} &=&-\frac{d_{0}\left( \varepsilon -1\right) ab\left(
a+b\right) \mathbf{e}_{y}}{\sqrt{b^{2}+\xi ^{\prime }}\left( \sqrt{a^{2}+\xi
^{\prime }}+\sqrt{b^{2}+\xi ^{\prime }}\right) \left( \varepsilon a+b\right)
}  \notag \\
&&+\frac{d_{0}\left( {\varepsilon -1}\right) ab\left( a+b\right) y^{\prime }%
}{\sqrt{a^{2}+\xi ^{\prime }}\left( b^{2}+\xi ^{\prime }\right) ^{3/2}}%
\left( \frac{x^{\prime }{}^{2}}{\left( a^{2}+\xi ^{\prime }\right) ^{2}}+%
\frac{y^{\prime }{}^{2}}{\left( b^{2}+\xi ^{\prime }\right) ^{2}}\right)
^{-1}  \notag \\
&&\times \left\{ \frac{x^{\prime }\mathbf{e}_{x}}{\left( a^{2}+\xi ^{\prime
}\right) \left( a+\varepsilon b\right) }+\frac{y^{\prime }\mathbf{e}_{y}}{%
\left( b+\xi ^{\prime }\right) \left( \varepsilon a+b\right) }\right\} .
\label{eq42}
\end{eqnarray}

C) Dipole moment is oriented along $z$-axis ($\mathbf{d}_{0}=d_{0}\mathbf{e}%
_{z}$ )

\begin{equation}
\delta \mathbf{d}=0  \label{eq43}
\end{equation}

Note that the expressions (\ref{eq41}) and (\ref{eq42}) in case of circular
nanocylinder ($a=b$) are significantly simplified and take the well known
form \cite{ref35}

\begin{equation}
\delta \mathbf{d}=\left( \frac{{\varepsilon -1}}{{\varepsilon +1}}\right)
\left( \frac{a}{\rho ^{\prime }}\right) ^{2}\left[ -\mathbf{d}_{0}+2\left(
\mathbf{d}_{0}\mathbf{n}^{\prime }\right) \mathbf{n}^{\prime }\right]
\label{eq44}
\end{equation}

\noindent where $\mathbf{d}_{0}=d_{0,x}\mathbf{e}_{x}+d_{0,y}\mathbf{e}_{y}$%
; $\mathbf{n}^{\prime }=\left( x^{\prime }{\mathbf{e}}_{x}{+}y^{\prime }{%
\mathbf{e}}_{y}\right) /\rho ^{\prime }$ and $\rho ^{\prime }=\sqrt{%
x^{\prime }{}^{2}+y^{\prime }{}^{2}}$.

Because the induced dipole moment (\ref{eq41})-(\ref{eq43}) is known, one
can find the radiative rate of the spontaneous decay by using Eq.(\ref{eq5}%
). Let an atom be placed on $x$-axis at point $x^{\prime }$. Then it is
sufficient to consider two orientations of the dipole moment, in accordance
with Eqs. (\ref{eq41})-(\ref{eq43}).

A) Dipole moment is oriented along $x$-axis

\begin{equation}
\left( \frac{\gamma }{\gamma _{0}}\right) ^{radiative}=\left\vert 1+\frac{%
ab\left( a+b\right) }{\sqrt{x^{\prime }{}^{2}-a^{2}+b^{2}}\left( x^{\prime }+%
\sqrt{x^{\prime }{}^{2}-a^{2}+b^{2}}\right) }\left( \frac{\varepsilon -1}{%
a+\varepsilon b}\right) \right\vert ^{2}.  \label{eq45}
\end{equation}

B) Dipole moment is oriented along $y$-axis

\begin{equation}
\left( \frac{\gamma }{\gamma _{0}}\right) ^{radiative}=\left\vert 1-\frac{%
ab\left( a+b\right) }{\sqrt{x^{\prime }{}^{2}-a^{2}+b^{2}}\left( x^{\prime }+%
\sqrt{x^{\prime }{}^{2}-a^{2}+b^{2}}\right) }\left( \frac{\varepsilon -1}{%
\varepsilon a+b}\right) \right\vert ^{2}.  \label{eq46}
\end{equation}

If the dipole moment of an atom is oriented along the nanocylinder axis ($z$%
-axis) then the radiative rate of the spontaneous decay should not be
changed.

If an atom is inside a dielectric elliptic cylinder without losses, then the
expression for the radiative rate of spontaneous decays will take the
following form at arbitrary dipole moment orientation, $\mathbf{d}%
_{0}=d_{0,x}\mathbf{e}_{x}+d_{0,y}\mathbf{e}_{y}+d_{0,z}\mathbf{e}_{z}$,

\begin{equation}
\left( \frac{\gamma }{\gamma _{0}}\right) ^{radiative}=\left( \frac{a+b}{%
a+\varepsilon b}\right) ^{2}\frac{d_{0,x}^{2}}{d_{0}^{2}}+\left( \frac{a+b}{%
\varepsilon a+b}\right) ^{2}\frac{d_{0,y}^{2}}{d_{0}^{2}}+\frac{d_{0,z}^{2}}{%
d_{0}^{2}},  \label{eq47}
\end{equation}

\noindent which is independent of the atomic position inside the
nanocylinder. In deriving (\ref{eq47}) we neglect the local field factor
\cite{ref55}. It should be noted that formula (\ref{eq47}) in agreement with
that obtained in \cite{ref56} for the case of two-dimensional ellipse (to
obtain that result we must substitute in (\ref{eq47}) $d_{0,z}=0$). It also
be noted, that expression (\ref{eq47}) can be applied to describe of an atom
in elliptic cylindrical cavity in material with permittivity $\varepsilon
_{0}$. In this case one should make substitution $\varepsilon \rightarrow
1/\varepsilon _{0}$ and multiply the obtained expression by $\sqrt{%
\varepsilon _{0}}$.

Note that Eqs. (\ref{eq45}) and (\ref{eq46}) have been derived as a limiting
case of a finite-size ellipsoid with fixed permittivity. These formulae are
unsuitable for describing the radiative decay rate of atoms near conducting
(metallic) cylinder, because the waveguided modes (plasmon modes) of the
infinite cylinder had not been taken into account. The correct expression
for the case of perfectly conducting cylinder is given in Eq.(\ref{eq35}).

\subsection{The rate of nonradiative decays}

In accordance with section 2, the rate of nonradiative decays of an atom
near a nanobody is determined by a quasistatic solution of the problem near
an elliptic nanocylinder. A Green's function of the Laplace equation
expressed in elliptical coordinates may be obtained from the Green's
function of the Helmholtz equation (the first terms in the right-hand side
of Eqs. (\ref{eq8}) and (\ref{eq9})) in the limit $k\rightarrow 0$. As a
result of such a procedure in the region $u_{0}<u<u^{\prime }$ the
expression for the scalar Green's function may be written in the form

\begin{eqnarray}
G_{0}\left( \mathbf{r,r}^{\prime }\right) &=&\frac{{1}}{{{\left\vert \mathbf{%
r-r}^{\prime }\right\vert }}}  \notag \\
&=&4i\sum\limits_{\sigma =e,o}\sum\limits_{n=0}^{\infty
}\int\limits_{0}^{\infty }d\alpha \cos \left( \alpha \left( z-z^{\prime
}\right) \right) \frac{S\sigma _{n}\left( i\alpha f,v\right) S\sigma
_{n}\left( i\alpha f,v^{\prime }\right) }{M\sigma _{n}\left( i\alpha
f\right) }  \notag \\
&&\times J\sigma _{n}\left( i\alpha f,u\right) H^{\left( {1}\right) }\sigma
_{n}\left( i\alpha f,u^{\prime }\right)  \label{eq48}
\end{eqnarray}

It should be noted here that function $S\sigma _{n}\left( i\alpha f,v\right)
$, as follows from (\ref{eq10}) and (\ref{eq11}), is the real function
because the coefficients $B\sigma _{n}\left( i\alpha f\right) $ are real.
Functions $J\sigma _{n}\left( i\alpha f,u\right) $ and $H^{\left( {1}\right)
}\sigma _{n}\left( i\alpha f,u\right) $ are either real or imaginary
depending on the parity of $n$. The product $J\sigma _{n}\left( i\alpha
f,u\right) H^{\left( {1}\right) }\sigma _{n}\left( i\alpha f,u^{\prime
}\right) $ is always purely imaginary. Hence, the right-hand side of Eq.(\ref%
{eq48}) is the real function.

Using Eq.(\ref{eq48}) one can obtain a quasistatic solution for the
potential of a dielectric nanocylinder, with dielectric constant $%
\varepsilon $, excited by the point-like dipole source located at the point
with ($u^{\prime },v^{\prime },z$) coordinates. The solution for the
potential $\phi ^{r}$ induced in space near the cylinder has the form:

\begin{eqnarray}
\phi ^{r} &=&4i\left( \mathbf{d}{_{0}{\nabla }^{\prime }}\right)
\sum\limits_{\sigma =e,o}\sum\limits_{n=0}^{\infty }\int\limits_{0}^{\infty
}d\alpha \cos \left( \alpha \left( z-z^{\prime }\right) \right) \frac{%
S\sigma _{n}\left( i\alpha f,v\right) S\sigma _{n}\left( i\alpha f,v^{\prime
}\right) }{M\sigma _{n}\left( i\alpha f\right) }  \notag \\
&&\times H^{\left( {1}\right) }\sigma _{n}\left( i\alpha f,u\right)
H^{\left( {1}\right) }\sigma _{n}\left( i\alpha f,u^{\prime }\right) A\sigma
_{n}\left( i\alpha f,u_{0}\right) ,  \label{eq50}
\end{eqnarray}

\noindent where the elliptical wave reflection coefficient has the form

\begin{eqnarray}
A\sigma _{n}\left( s,u_{0}\right) &=&\left( \varepsilon -1\right)
\left. \frac{{\partial }}{\partial w}J\sigma _{n}\left( s,w\right)
\right\vert _{w=u_{0}}J\sigma _{n}\left( s,u_{0}\right)  \notag \\
&&\times \left\{ \left. \frac{{\partial }}{\partial w}H^{\left(
{1}\right) }\sigma _{n}\left( s,w\right) \right\vert
_{w=u_{0}}J\sigma _{n}\left( s,u_{0}\right) \right.  \notag \\
&&\left. -\varepsilon \left. \frac{{\partial }}{\partial w}J\sigma
_{n}\left( s,w\right) \right\vert _{w=u_{0}}H^{\left( {1}\right)
}\sigma _{n}\left( s,u_{0}\right) \right\}^{-1} .  \label{eq51}
\end{eqnarray}

For metallic elliptic cylinders with low losses ($\varepsilon ={\varepsilon }%
^{\prime }+i{\varepsilon }^{\prime \prime }$; ${\varepsilon }^{\prime }<0$, $%
{\varepsilon }^{\prime \prime }\ll {\left\vert {{\varepsilon }^{\prime }}%
\right\vert }$) denominator of (\ref{eq51}) becomes close to zero for some
longitudinal wavenumbers (plasmon modes). In this case quasistatic
approximation is not valid and one should use full set of Maxwell's
equations. In the case of elliptic nanocylinder made of real metals ($%
\varepsilon ={\varepsilon }^{\prime }+i{\varepsilon }^{\prime \prime }$, ${%
\varepsilon }^{\prime \prime }\sim 1$) or dielectrics the denominator of (%
\ref{eq51}) is always nonzero and quasistatic approximation is valid in this
case. So we restrict ourselves for the case of dielectric elliptic
nanocylinders made of real metals or dielectrics.

By calculating the reflected field $\mathbf{E}^{r}=-\nabla \phi ^{r}$ with
the help of (\ref{eq50}) and using (\ref{eq2}) and (\ref{eq4})we obtain the
following expressions for the quasistatic contribution to the nonradiative
rate of the spontaneous decay of an atom near the dielectric (metallic)
nanocylinder.

\begin{center}
\textit{Dipole moment is oriented along x-axis (}$\mathbf{d}_{0}=d_{0}%
\mathbf{e}_{x}$\textit{)}

\begin{eqnarray}
\left( \frac{\gamma }{\gamma _{0}}\right) _{x}^{nonrad} &=&-\frac{{6}}{%
k^{3}f^{2}}\sum\limits_{\sigma =e,o}\sum\limits_{n=0}^{\infty
}\int\limits_{0}^{\infty }d\alpha \frac{\left[ S\sigma _{n}\left( i\alpha
f,v^{\prime }\right) \right] ^{2}}{M\sigma _{n}\left( i\alpha f\right) }%
\left[ H^{\left( {1}\right) }\sigma _{n}\left( i\alpha f,u^{\prime }\right) %
\right] ^{2}  \notag \\
&&\times \left\{ v^{\prime }\left( \frac{u^{\prime }{}^{2}-1}{u^{\prime
}{}^{2}-v^{\prime }{}^{2}}\right) \frac{\left. \left( {\partial /}\partial
w\right) H^{\left( {1}\right) }\sigma _{n}\left( i\alpha f{,}w\right)
\right\vert _{w=u^{\prime }}}{H^{\left( {1}\right) }\sigma _{n}\left(
i\alpha f,u^{\prime }\right) }\right.  \notag \\
&&\left. +u^{\prime }\left( \frac{1-v^{\prime }{}^{2}}{u^{\prime
}{}^{2}-v^{\prime }{}^{2}}\right) \frac{\left. \left( {\partial /}\partial
w\right) S\sigma _{n}\left( i\alpha f,w\right) \right\vert _{w=v^{\prime }}}{%
S\sigma _{n}\left( i\alpha f,v^{\prime }\right) }\right\} ^{2}  \notag \\
&&\times \text{Re}\left[ A\sigma _{n}\left( i\alpha f,u_{0}\right) \right] .
\label{eq52}
\end{eqnarray}

\textit{Dipole moment is oriented along y-axis (}$\mathbf{d}_{0}=d_{0}%
\mathbf{e}_{y}$\textit{\ )}

\begin{eqnarray}
\left( \frac{\gamma }{\gamma _{0}}\right) _{y}^{nonrad} &=&-\frac{{6}\left(
u^{\prime }{}^{2}-1\right) \left( 1-v^{\prime }{}^{2}\right) }{%
k^{3}f^{2}\left( u^{\prime }{}^{2}-v^{\prime }{}^{2}\right) ^{2}}%
\sum\limits_{\sigma =e,o}\sum\limits_{n=0}^{\infty }\int\limits_{0}^{\infty
}d\alpha \frac{\left[ S\sigma _{n}\left( i\alpha f,v^{\prime }\right) \right]
^{2}}{M\sigma _{n}\left( i\alpha f\right) }  \notag \\
&&\times \left[ H^{\left( {1}\right) }\sigma _{n}\left( i\alpha f,u^{\prime
}\right) \right] ^{2}\left\{ u^{\prime }\frac{\left. \left( {\partial /}%
\partial w\right) H^{\left( {1}\right) }\sigma _{n}\left( i\alpha f,w\right)
\right\vert _{w=u^{\prime }}}{H^{\left( {1}\right) }\sigma _{n}\left(
i\alpha f,u^{\prime }\right) }\right.  \notag \\
&&\left. -v^{\prime }\frac{\left. \left( {\partial /}\partial w\right)
S\sigma _{n}\left( i\alpha f,w\right) \right\vert _{w=v^{\prime }}}{S\sigma
_{n}\left( i\alpha f,v^{\prime }\right) }\right\} ^{2}\text{Re}\left[
A\sigma _{n}\left( i\alpha f,u_{0}\right) \right] .  \label{eq53}
\end{eqnarray}

\textit{Dipole moment is oriented along z-axis (}$\mathbf{d}_{0}=d_{0}%
\mathbf{e}_{z}$\textit{\ )}

\begin{eqnarray}
\left( \frac{\gamma }{\gamma _{0}}\right) _{z}^{nonrad} &=&-\frac{{6}}{k^{3}}%
\sum\limits_{\sigma =e,o}\sum\limits_{n=0}^{\infty }\int\limits_{0}^{\infty
}d\alpha \alpha ^{2}\frac{\left[ S\sigma _{n}\left( i\alpha f,v^{\prime
}\right) \right] ^{2}}{M\sigma _{n}\left( i\alpha f\right) }  \notag \\
&&-\frac{{6}}{k^{3}}\sum\limits_{\sigma =e,o}\sum\limits_{n=0}^{\infty
}\int\limits_{0}^{\infty }d\alpha \alpha ^{2}\frac{\left[ S\sigma _{n}\left(
i\alpha f,v^{\prime }\right) \right] ^{2}}{M\sigma _{n}\left( i\alpha
f\right) }.  \label{eq54}
\end{eqnarray}
\end{center}

If an atom is placed very close to the nanocylinder the nonradiative
spontaneous decay rate may be approximated by the expressions, which
describe nonradiative processes near a plane surface \cite{ref58}. For
dipole with tangential and normal to surface of cylinder orientations we
have respectively

\begin{eqnarray}
\left( \frac{\gamma }{\gamma _{0}}\right) _{tang}^{nonrad} &=&\frac{{3}}{{16}%
\left( k\Delta ^{\prime }\right) ^{3}}\text{Im}\left( \frac{\varepsilon -1}{%
\varepsilon +1}\right) ,  \notag \\
\left( \frac{\gamma }{\gamma _{0}}\right) _{norm}^{nonrad} &=&2\left( \frac{%
\gamma }{\gamma _{0}}\right) _{tang}^{nonrad}.  \label{eq55}
\end{eqnarray}

\noindent Here ${\Delta }^{\prime }$ is the distance from atom to
surface; in the case
of an elliptic cylinder we can write $\Delta ^{\prime }=\zeta ^{\prime }%
\sqrt{\frac{x^{\prime }{}^{2}}{\left( \zeta ^{\prime }+a^{2}\right) ^{2}}+%
\frac{y^{\prime }{}^{2}}{\left( \zeta ^{\prime }+b^{2}\right) ^{2}}}$, where
$\zeta ^{\prime }$ is the positive root of the equation $\frac{x^{\prime
}{}^{2}a^{2}}{\left( \zeta ^{\prime }+a^{2}\right) ^{2}}+\frac{y^{\prime
}{}^{2}b^{2}}{\left( \zeta ^{\prime }+b^{2}\right) ^{2}}=1$.

\section{Graphic illustrations and discussion of results}

Let us first consider an atom near a dielectric nanocylinder. Figure 4
illustrates the radiative rate of the spontaneous decay (\ref{eq45}) and (%
\ref{eq46}) near the dielectric nanocylinder. As an atom moves away from the
surface the decay rate tends to the value, which is corresponds to the case
of an atom in free space in the absence of a nanobody. As the dielectric
constant of the nanocylinder material is increased, the radiative rate for
the normally oriented dipole moment of an atom increases by tending to large
(finite) value, and for the tangentially oriented dipole moment it decreases
by tending to low values (zero). The most increase in the decay rate occurs
for the normally-oriented atom in the direction of the larger semi-axis
(Fig.4(a)).

From Fig.5 it is seen how the decay rate changes for an atom located in
close proximity to the surface of a nanocylinder depending on the ratio
between the $a$ and $b$ elliptic semi-axes. If one of the nanocylinder
semi-axes tends to zero (the case of a dielectric nanobelt) then from (\ref%
{eq45}) and (\ref{eq46}) it follows that $\gamma ^{radiative} / \gamma _{0}
\to \varepsilon ^{2}$ for the normally-oriented dipole moment of an atom
placed at the belt edge, while for the tangential orientation of the dipole
moment we obtain $\gamma ^{radiative} / \gamma _{0} \to 1 / \varepsilon ^{2}$
(Fig.5(a)). For the atom placed at the belt facet, independent of the dipole
moment orientation, we obtain $\gamma ^{radiative} / \gamma _{0} \to 1$
(Fig.5(b)).

Figure 6 illustrates radiative and nonradiative rates of the spontaneous
decay as the function of the distance of an atom to the surface of the
dielectric elliptic nanocylinder having the permittivity $\varepsilon
=3+i10^{-8}$. The atom is shifted along the $x$-axis and its dipole moment
is oriented along this axis. To calculate the Mathieu functions we have used
of the method \cite{ref59} based on the solution of the generalized
eigen-value problem for the coefficients $B\sigma _{n}^{m}$. The integration
of (\ref{eq52})-(\ref{eq54}) was performed by the mean-value method using
the Aitken's process \cite{ref60} for acceleration of the integral
convergence. As seen from Fig.6 the radiative rate of the spontaneous decay
for a dielectric with a very small imaginary part of permittivity is
dominating, as a whole, over a nonradiative rate, and takes maximal values
at the cylinder surface. But if the atom is placed very close to the
nanocylinder surface, the nonradiative rate exceeds over the radiative one,
and at $u^{\prime }=u_{0}$ it turns to the infinity, as follows from the
asymptotic expression (\ref{eq55}) depicted by dotted line in Fig. 6. Let us
determine the distance, on which the radiative and nonradiative rates for
the normally oriented dipole moment of an atom coincide. By equating (\ref%
{eq45}) with (\ref{eq55}), for the case of a small imaginary part of the
permittivity $\varepsilon ={\varepsilon }^{\prime }+i{\varepsilon }^{\prime
\prime }$ of a cylinder (${\varepsilon }^{\prime \prime }\ll {\varepsilon }%
^{\prime }$), we get that at

\begin{equation}
\frac{x^{\prime }-a}{a}\approx \frac{{1}}{ka}\left\{ {{\frac{{3}}{{4}}}{%
\frac{\varepsilon ^{\prime \prime }}{{\left( {1+{\varepsilon }^{\prime }}%
\right) ^{2}}}}}\frac{\left( a+\varepsilon ^{\prime }b\right) ^{2}}{%
\varepsilon ^{\prime }{}^{2}\left( a+b\right) ^{2}}\right\} ^{1/3}
\label{eq56}
\end{equation}

\noindent the nonradiative contribution to decay rate is equal to the
radiative one. For the case represented in Fig.6 we obtain: $\left(
x^{\prime }-a\right) /a\approx 4.4\times 10^{-3}$, which means that one may
actually neglect here the nonradiative processes. If the atom is $y$-shifted
and has the dipole moment oriented along this axis, then from (\ref{eq56})
we obtain

\begin{equation}
\frac{y^{\prime }-b}{a}\approx \frac{{1}}{ka}\left\{ {{\frac{{3}}{{4}}}{%
\frac{{{\varepsilon }^{\prime \prime }}}{{\left( {{\varepsilon }^{\prime }+1}%
\right) ^{2}}}}}\frac{\left( \varepsilon ^{\prime }a+b\right) ^{2}}{%
\varepsilon ^{\prime }{}^{2}\left( a+b\right) ^{2}}\right\} ^{1/3}.
\label{eq57}
\end{equation}

Figure 7 illustrates decay rates for a silver nanocylinder with $\varepsilon
=-3.02+i0.21$ ($\lambda =373$ nm \cite{ref61}). The ratio of the elliptical
semi-axes is $a$/$b$=3 and $a$=6 nm ($ka\approx 0.1)$. As is well seen, the
region nearby cylinder, where the nonradiative rate is greater than
radiative, is essentially wider that the region represented in Fig.6. This
is due to a larger imaginary part of permittivity of the nanocylinder
material. At small distances from the cylinder surface the nonradiative rate
of the spontaneous decay increases in accordance with (\ref{eq55}) (dotted
line of Fig.7), and at large distances from the cylinder surface, it tends
to zero.

Now let us consider the case of a perfectly conducting cylinder. The
integration of Eqs. (\ref{eq17})-(\ref{eq26}) was performed by the method of
mean values with using the Aitken's process. Figure 8 illustrates the
spontaneous decay rate of an atom placed near a perfectly conducting
elliptic cylinder ($ka=1$), as the function of the atom's position along the
larger semi-axis $a$. As is seen, the decay rate achieves the maximum in
close proximity to the cylinder surface ($u^{\prime }=u_{0})$ for the
normally oriented dipole moment of an atom (x-line of Fig.7(a) or y-line of
Fig.8(b)). For the tangentially oriented dipole moment we obtain zero at the
surface of the cylinder (y-line of Fig.8(a) or x-line of Fig.8(b), and also
z--line) resulting from a full compensation of the dipole moment of an atom.
As the ratio of semi-axes increases (when the smaller semi-axis $b$
decreases) the maximum rate of the spontaneous decay is increased (cf.
x-lines in Fig.8(a) and Fig.8(b)) and tends to infinitely large values in
the limit of $b\rightarrow 0$, which agrees with the asymptotic formula (\ref%
{eq35}) for the nanocylinder.

Figure 9 illustrates the spontaneous decay rate of an atom placed near a
perfectly conducting elliptic cylinder ($ka=1$) as the function of the
atomic displacement along y-axis (towards a smaller semi-axis). The decay
rate achieves the maximal value (normal orientation of the dipole moment)
and the minimal value (tangential orientation) at the surface of a cylinder.
At high values of the semi-axes ratio, the curves y and z (Fig.9(b) for $%
a/b=100\gg 1$) look like the well known solutions \cite{ref58} for the
spontaneous decay rate of an atom near a perfectly conducting plane surface
with the normal (curve y) and tangential (curve z) orientations of the
dipole moment. The maximum value of the decay rate here is about 2. The
behavior of decay rate of an atom with x orientation of dipole momentum is
slightly different from the case of the plane mirror because of edge effects
in case of strip.

Figure 10 illustrates the decay rate of an atom placed in close proximity to
the surface of a perfectly conducting cylinder ($u^{\prime }=u_{0})$. The
dipole moment is oriented normally to the cylinder surface along the larger
semi-axis (Fig.10(a)) and along the smaller semi-axis (Fig.10(b)). As seen
from Fig.10, the expression (\ref{eq35}) obtained in the case of $%
ka\rightarrow 0$ for the atom with normal orientation, tends to infinitely
large values, which disagrees with the finite values derived from the
quasistatic expression (\ref{eq45}) in the limit of $\varepsilon \rightarrow
\infty $. This difference follows from the fact that the amplitude of the
surface current propagating along the cylinder at $ka\rightarrow 0$
increases infinitely large \cite{ref26}. In other words, perfectly
conducting cylinder of infinitely small cross-section is a kind of antenna,
which is effectively excited by a point dipole (atom). This means that the
induced dipole moment of the perfectly conducting cylinder and the rate of
spontaneous decay of an atom also tend to infinity at $ka\rightarrow 0$ (see
Eq.(\ref{eq36})). The analogous process is also observed in the case of a
circular nanocylinder (dotted line in Fig.10, Eq.(\ref{eq37})).

Figure 11 illustrates a comparison for the rate of spontaneous decay (\ref%
{eq16}) obtained by numerical integration (\ref{eq34}) and an asymptotic
expression (\ref{eq35}) for the atom with the normally oriented dipole
moment in close proximity to the surface ($u^{\prime }=u_{0})$ of perfectly
conducting cylinder. As it seen, the asymptotic expression agrees well with
the numerical data up to $ka=1$. It can, therefore, be suitable for a fast
evaluation of the spontaneous decay rate near the nanocylinder and for
verification of the numerical solution.

\section{Conclusion}

Thus, in this paper we obtained the analytical expressions for the radiative
rate of the spontaneous decay of an atom near a dielectric nanocylinder of
elliptical cross-section within the framework of a quasistatic
approximation. The retardation effects are taken into account for the
spontaneous decay of an atom near a perfectly conducting elliptic cylinder,
where the analytical solution can be found at arbitrary parameters of the
problem. Special attention is paid to the case of a nanocylinder where the
largest semi-axis of the cylinder is much less than wavelength. It is shown
that a decrease in the cross-section of a metallic (perfectly conducting)
nanocylinder leads to an infinite increase in the spontaneous decay rate for
a normally oriented dipole moment of the atom. A simple asymptotic
expression for the spontaneous decay rate has been obtained in the case of a
perfectly conducting nanocylinder. For the atom placed in close proximity to
a belt edge one can obtain arbitrary large values of the decay rate by means
of tuning elliptic semi-axes. The obtained results can be used for checking
the numerical solution in a more complex case of the dielectric elliptic
cylinder. The results of this work may be also helpful at studying an
interaction of atoms (molecules) with nanobelts and nanoribbons.

\bigskip \textbf{Acknowledgement.} The authors are grateful to the Russian
Fund of Fundamental Studies (grants 05-02-19647 and 07-02-01328) and the RAS
Presidium Program \textquotedblleft The effect of atomic-crystalline and
electron structures on properties of the condensed media\textquotedblright\
for a partial financial support of the present work.

One of the authors (D.V.G.) is thankful to the Young scientists support
Program of the Educational-Scientific Center of P.N. Lebedev Physical
Institute and the RAS Presidium Program \textquotedblleft Support of young
scientists\textquotedblright\ for a partial financial support of the present
work.

\pagebreak

\newpage

\begin{center}
\bigskip {\LARGE List of Figure Captions}
\end{center}

\bigskip

Fig.1 The nanobelts \cite{ref39}.

Fig.2 Geometry of the problem.

Fig.3 Elliptic cylinder coordinate system.

Fig.4 Radiative decay rate of an atom near a dielectric elliptic
nanocylinder as the function of the position of an atom. (a) atom is shifted
along $x$-axis; (b) atom is shifted along $y$-axis. Dipole moment
orientation: x - along $x$-axis; y - along $y$-axis. Solid line corresponds
to $\varepsilon =3$; dotted line corresponds to $\varepsilon =10$.
Nanocylinder semi-axes ratio $a/b=5$.

Fig.5 Radiative decay rate of an atom near the surface of a dielectric ($%
\varepsilon =3$) elliptic nanocylinder as the function of the semi-axes
ratio for the atom in close proximity to the surface of the nanocylinder.
(a) atom is placed on $x$-axis at point $x^{\prime }=a$; (b) atom is placed
on $y$-axis at point $y^{\prime }=a$. Dipole moment orientation: x - along $x
$-axis; y - along $y$-axis.

Fig.6 Radiative and nonradiative decay rates of an atom as the functions of
its position relative to a dielectric elliptic nanocylinder with
permittivity $\varepsilon =3+i10^{-8}$. The ratio of the nanocylinder
semi-axes $a/b=6$; $ka=0.1$. The atom is shifted by $x$-axis with the dipole
moment oriented along this axis. The dotted line corresponds to the
asymptotic form (\ref{eq55}).

Fig.7 Radiative and nonradiative decay rates of an atom as the functions of
its position relative to the silver elliptic nanocylinder with permittivity $%
\varepsilon =-3.02+i0.21$ ($\lambda =373$ nm \cite{ref61}). The atom is
shifted by $x$-axis with the dipole moment oriented along this axis. The
ratio of the nanocylinder semi-axes is $a/b=3$; $a=6$ nm. The dotted line
corresponds to the asymptotic form (\ref{eq55}).

Fig.8 Decay rate of an atom placed near a perfectly conducting
elliptic cylinder ($ka=1$) as the function of atom's position
relative to the cylinder. The atom is shifted along the $x$-axis.
Dipole moment orientation: x - along $x$-axis; y - along $y$-axis;
z - along $z$-axis. The semi-axes ratio: (a) $a/b=3$; (b)
$a/b=10$.

Fig.9 Decay rate of an atom placed near a perfectly conducting
elliptic cylinder ($ka=1$) as the function of atom's position
relative to the cylinder. The atom is shifted along the $y$-axis.
Dipole moment orientation: x - along $x$-axis; y - along $y$-axis;
z - along $z$-axis. The semi-axes ratio: (a) $a/b=3$; (b)
$a/b=100$.

Fig.10 Decay rate of an atom placed in close proximity to the surface of a
perfectly conducting elliptic cylinder ($u^{\prime }=u_{0}$ ), as the
function of $ka$. The ratio of the cylinder semi-axes: 1 - $a/b=3$; 2 - $%
a/b=10$; 3 - $a/b=100$. The dotted line corresponds to the case of $a/b=1$.
(a) atom is located on $x$-axis at point $x=x^{\prime }$, and its dipole
moment is oriented along $x$-axis; (b) atom is located on $y$-axis at point $%
y=y^{\prime }$, and its dipole moment is oriented along $y$-axis.

Fig.11 A comparison of an asymptotic expression for the spontaneous decay
rate (\ref{eq35}) with (\ref{eq16}) obtained by numerical integration of (%
\ref{eq34}) for the atom with the normal dipole moment orientation placed in
close proximity to the perfectly conducting cylinder surface ($u^{\prime
}=u_{0}$ ). A solid line corresponds to a numerical solution; the crosses
denote asymptotic expression. The ratio of semi-axes: 1 - $a/b=3$ and 2 - $%
a/b=1 $. Position of an atom relative to the cylinder surface is
marked by arrow.

\newpage
\begin{figure}
\centering\includegraphics[height=8cm,angle=0]{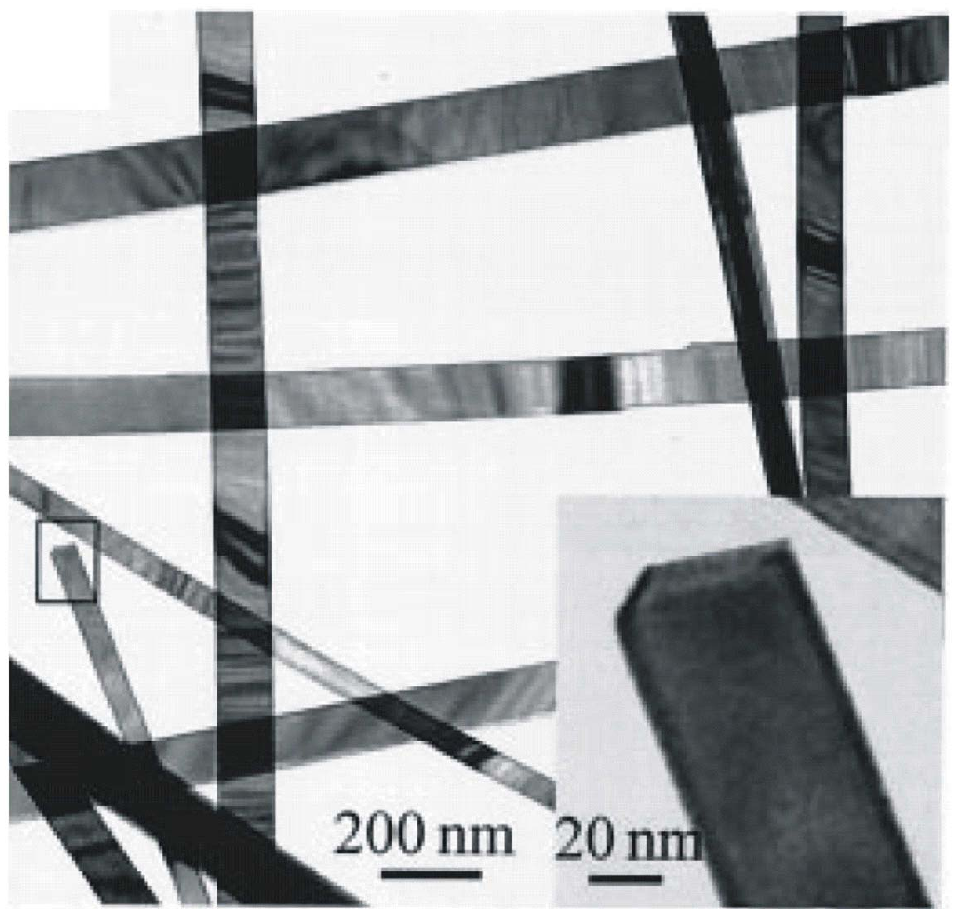}
\caption{}
\end{figure}
\pagebreak

\newpage
\begin{figure}
\centering\includegraphics[height=3cm,angle=0]{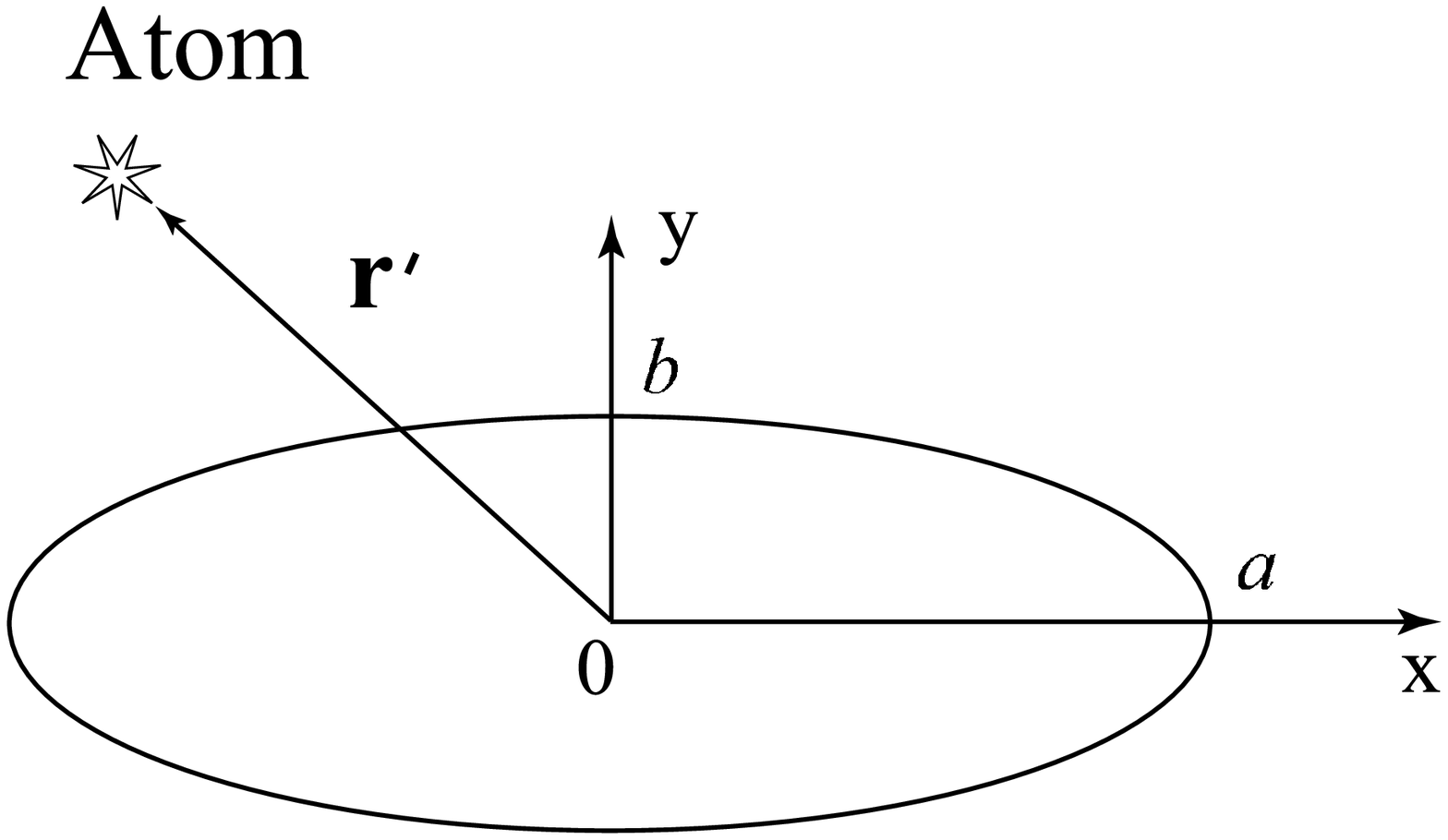}
\caption{}
\end{figure}
\pagebreak

\newpage
\begin{figure}
\centering \includegraphics[height=6cm,angle=0]{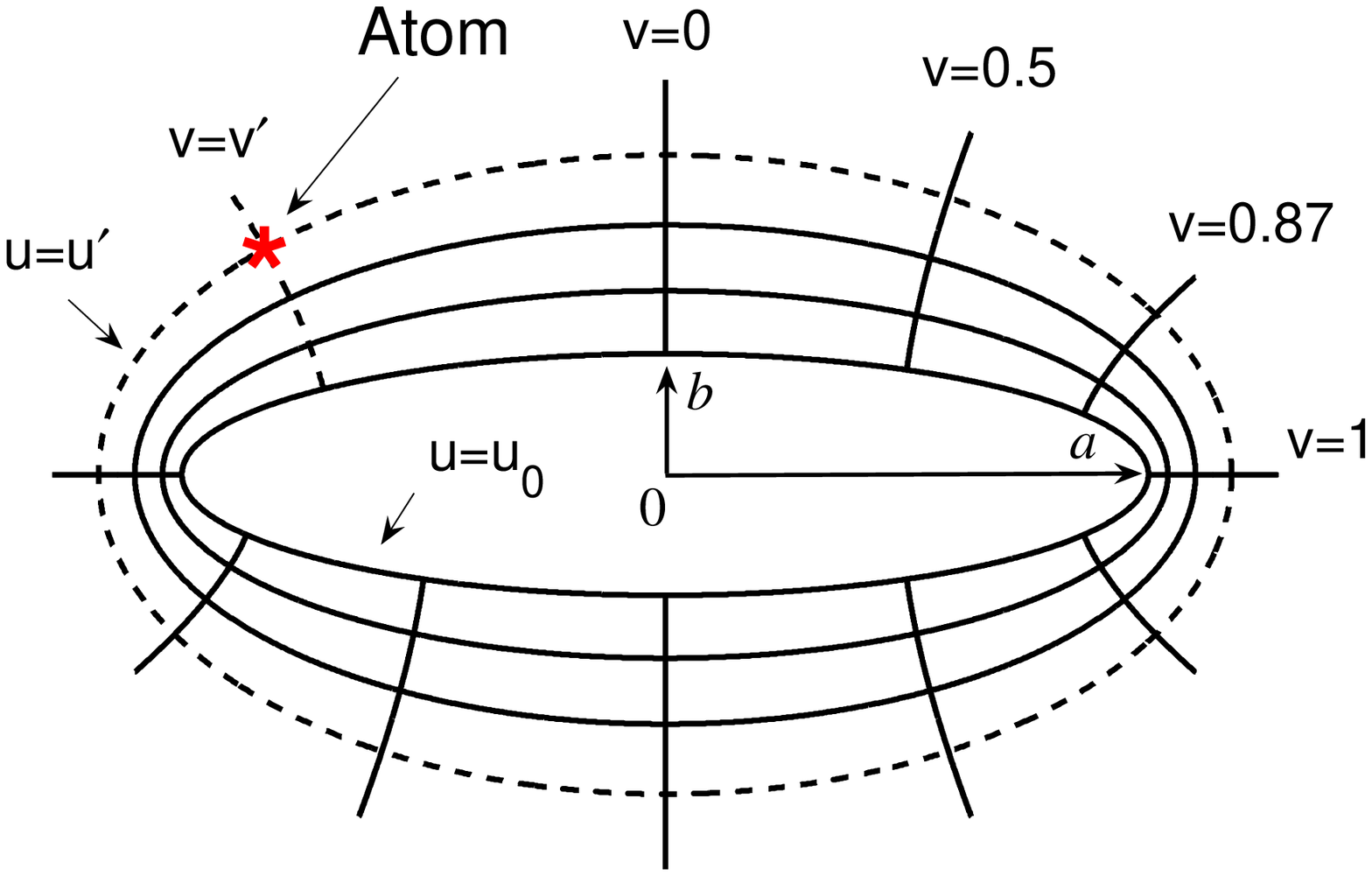}
\caption{}
\end{figure}
\pagebreak

\newpage
\begin{figure}
\centering \includegraphics[height=15cm,angle=0]{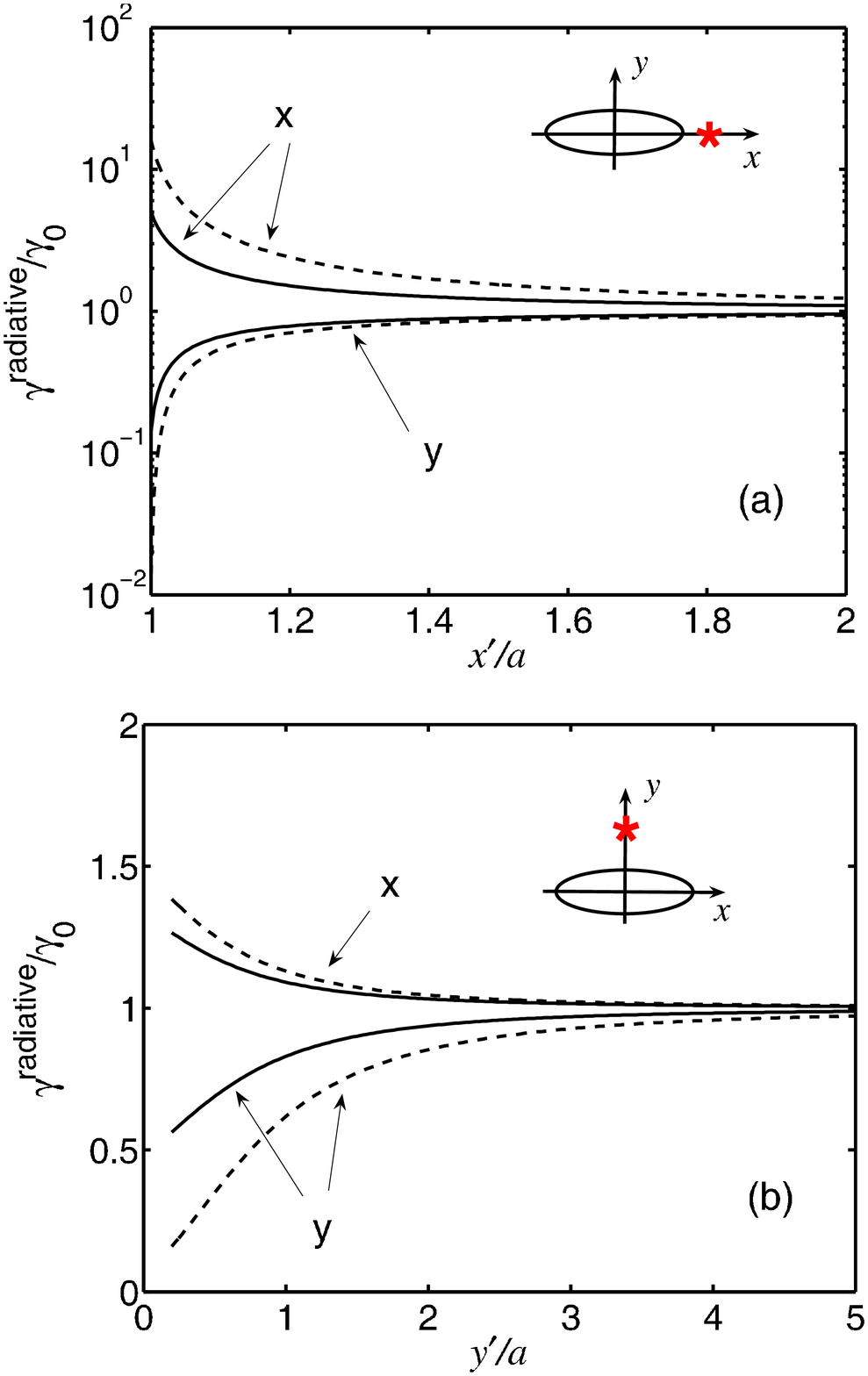}
\caption{}
\end{figure}
\pagebreak

\newpage
\begin{figure}
\centering \includegraphics[height=15cm,angle=0]{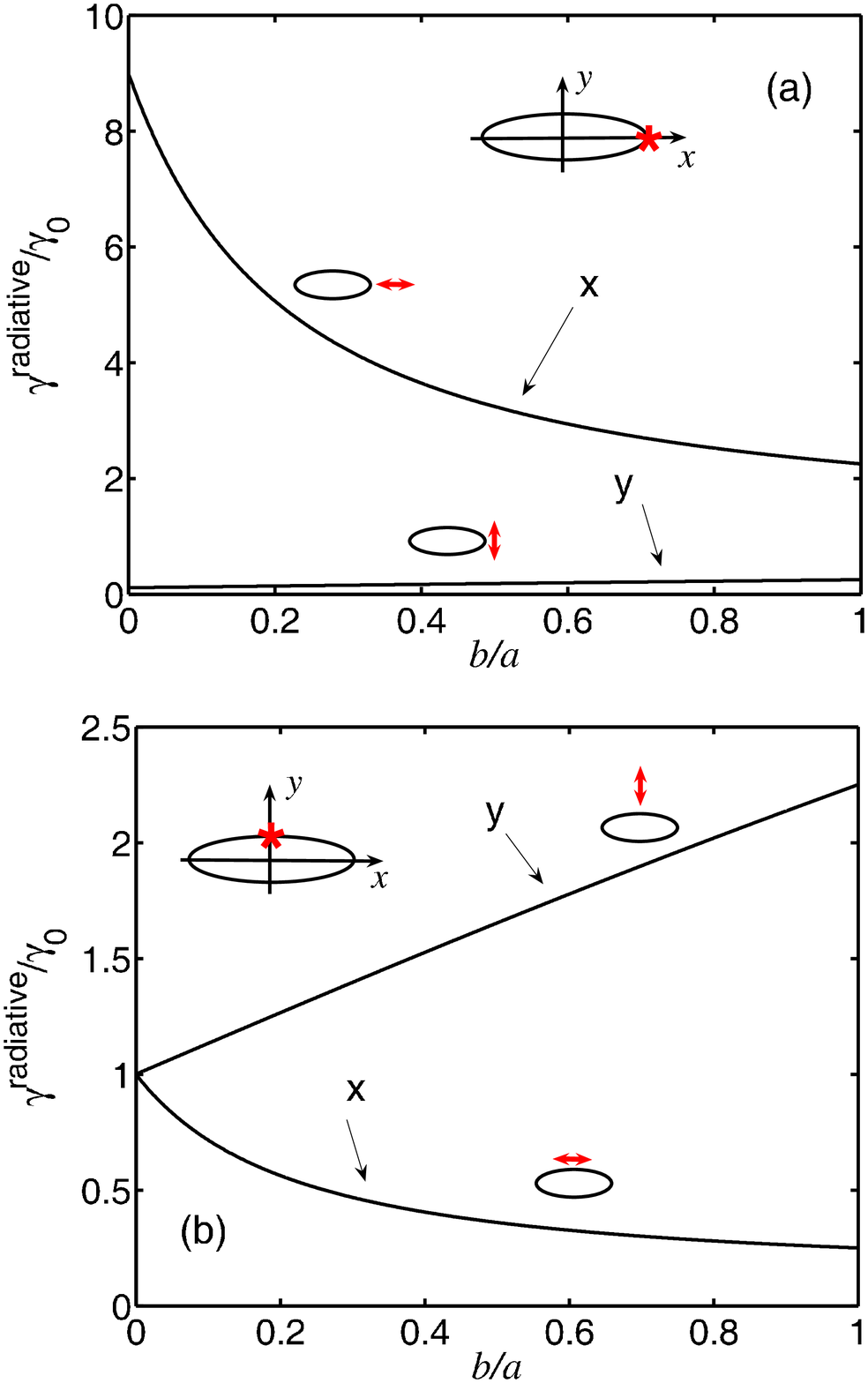}
\caption{}
\end{figure}
\pagebreak

\newpage
\begin{figure}
\centering \includegraphics[height=8cm,angle=0]{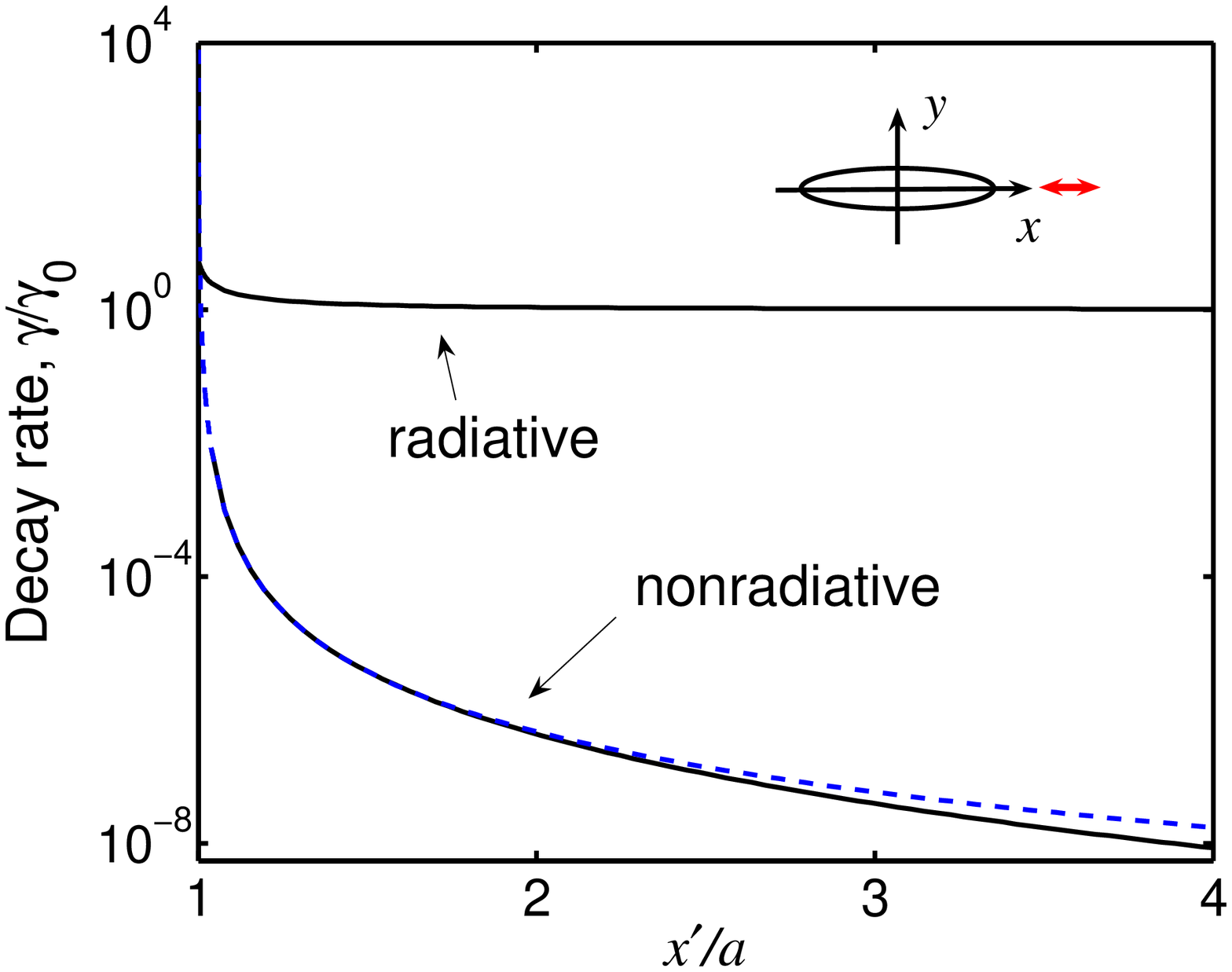}
\caption{}
\end{figure}
\pagebreak

\newpage
\begin{figure}
\centering \includegraphics[height=8cm,angle=0]{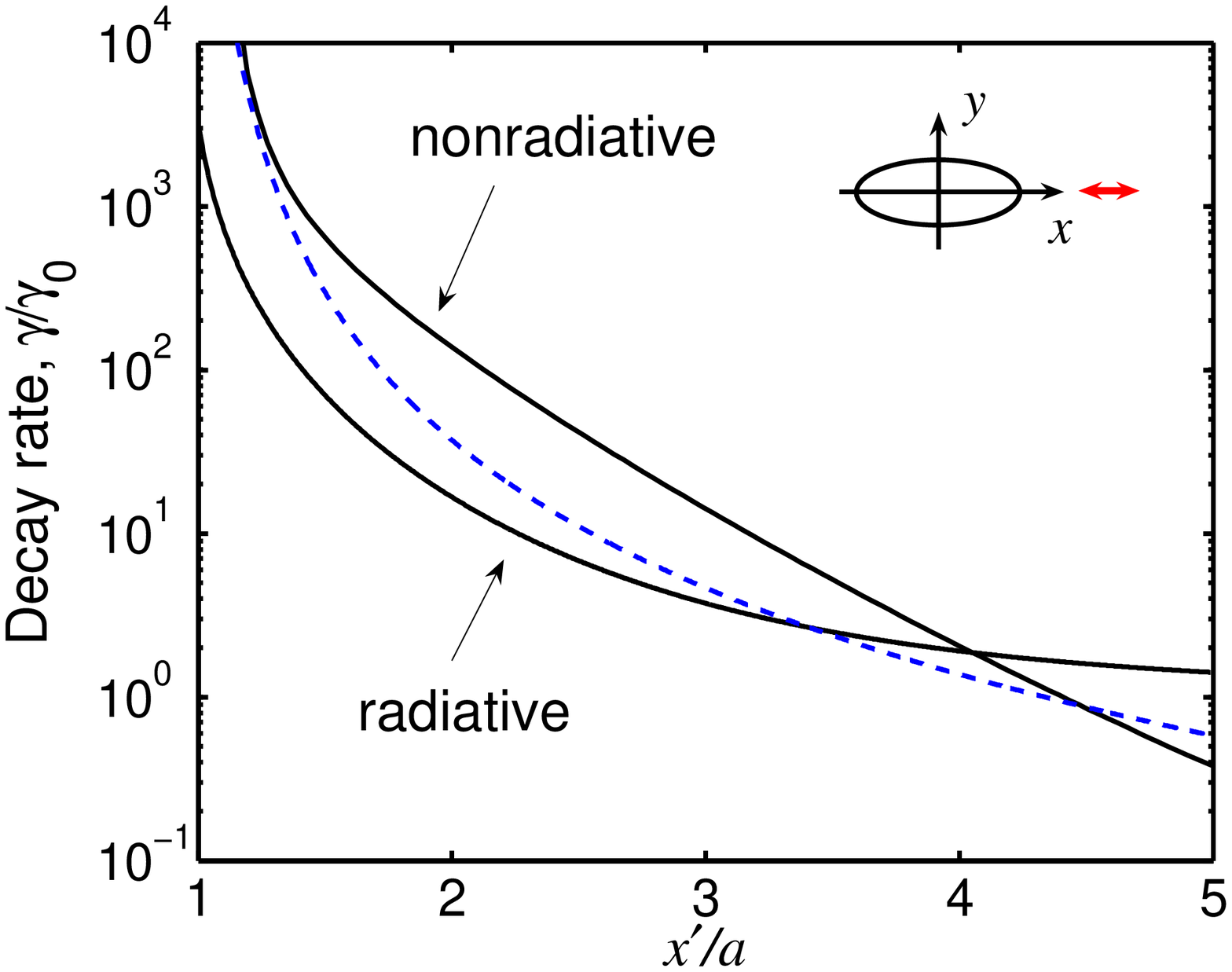}
\caption{}
\end{figure}
\pagebreak

\newpage
\begin{figure}
\centering \includegraphics[height=15cm,angle=0]{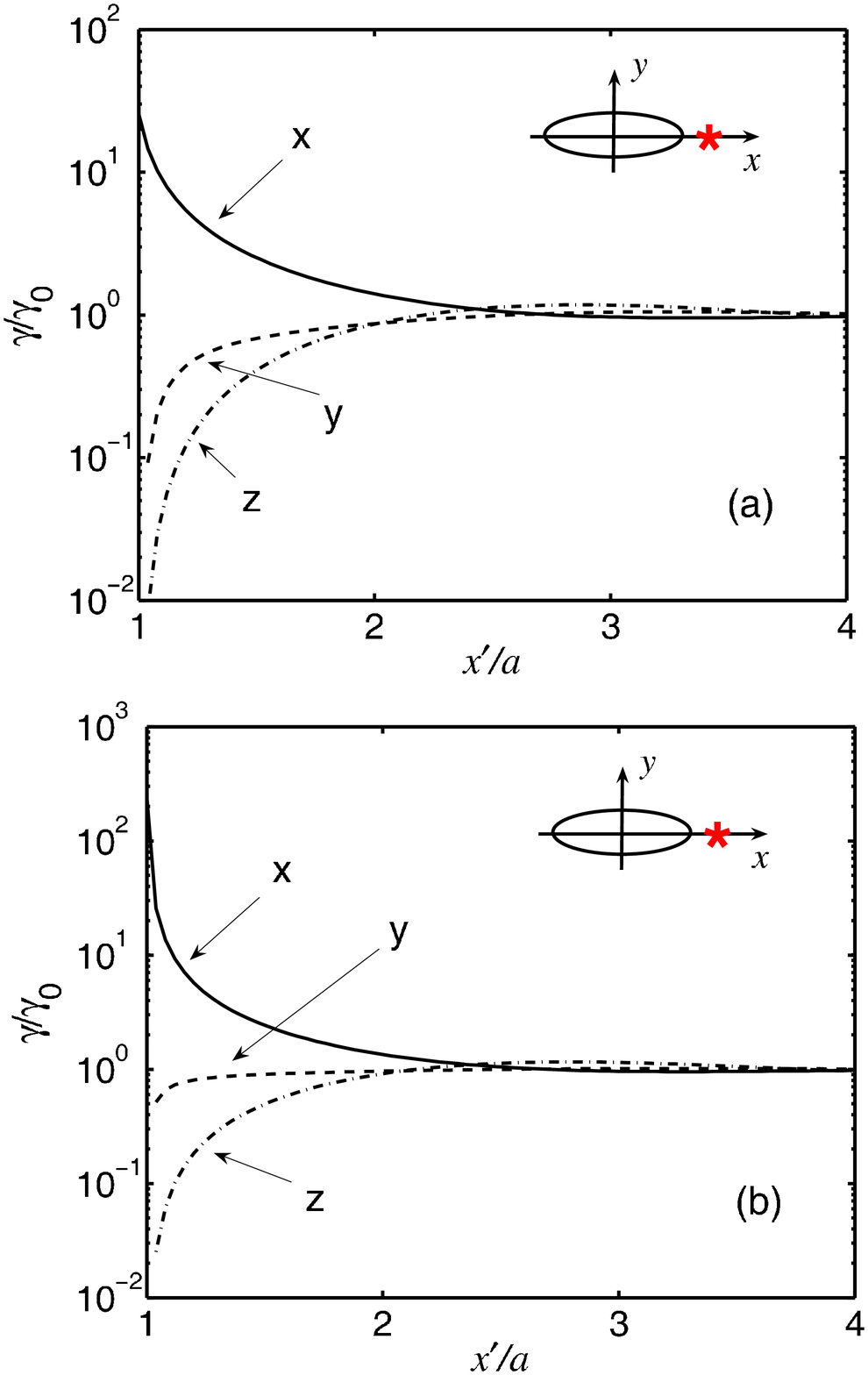}
\caption{}
\end{figure}
\pagebreak

\newpage
\begin{figure}
\centering \includegraphics[height=15cm,angle=0]{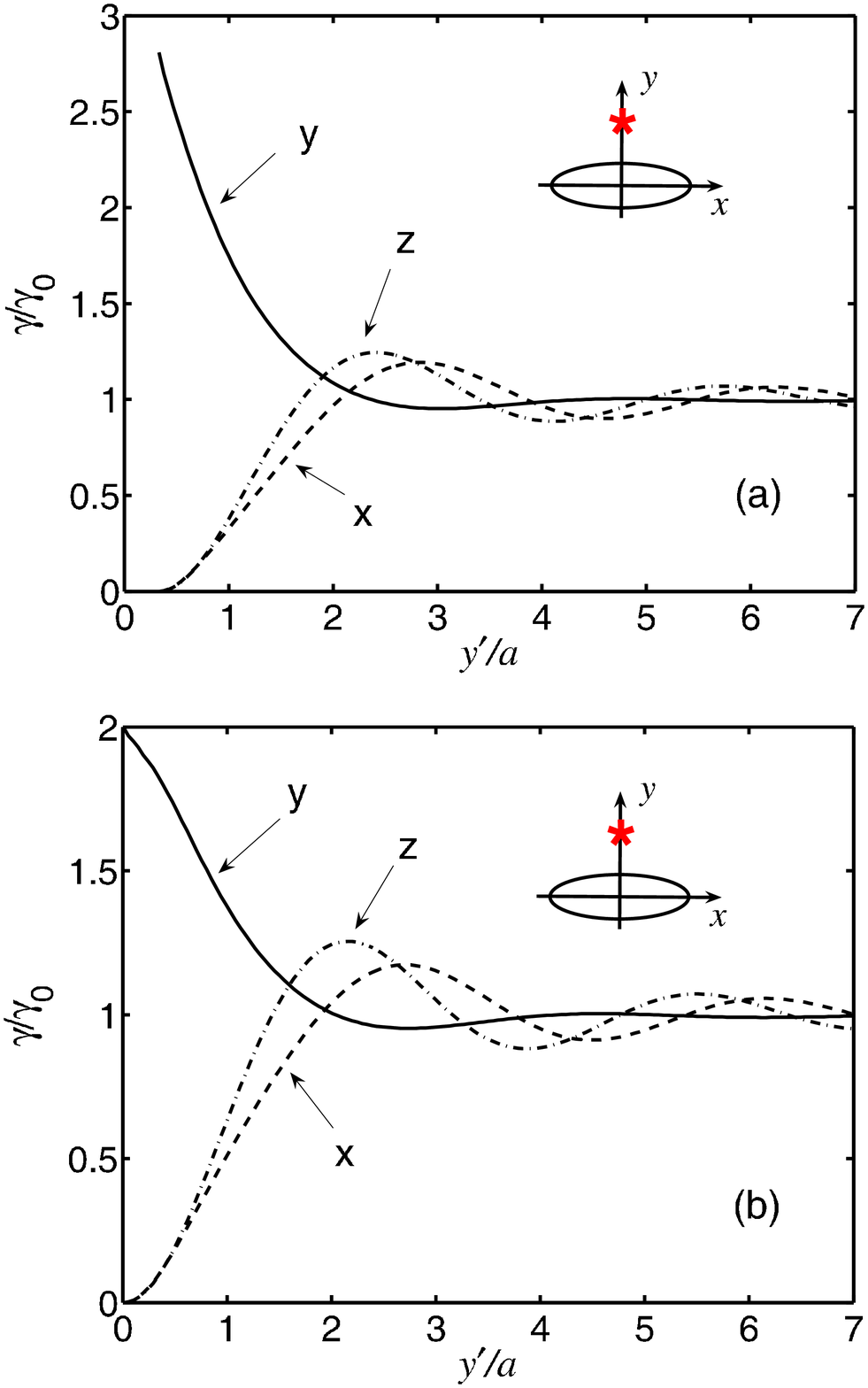}
\caption{}
\end{figure}
\pagebreak

\newpage
\begin{figure}
\centering \includegraphics[height=15cm,angle=0]{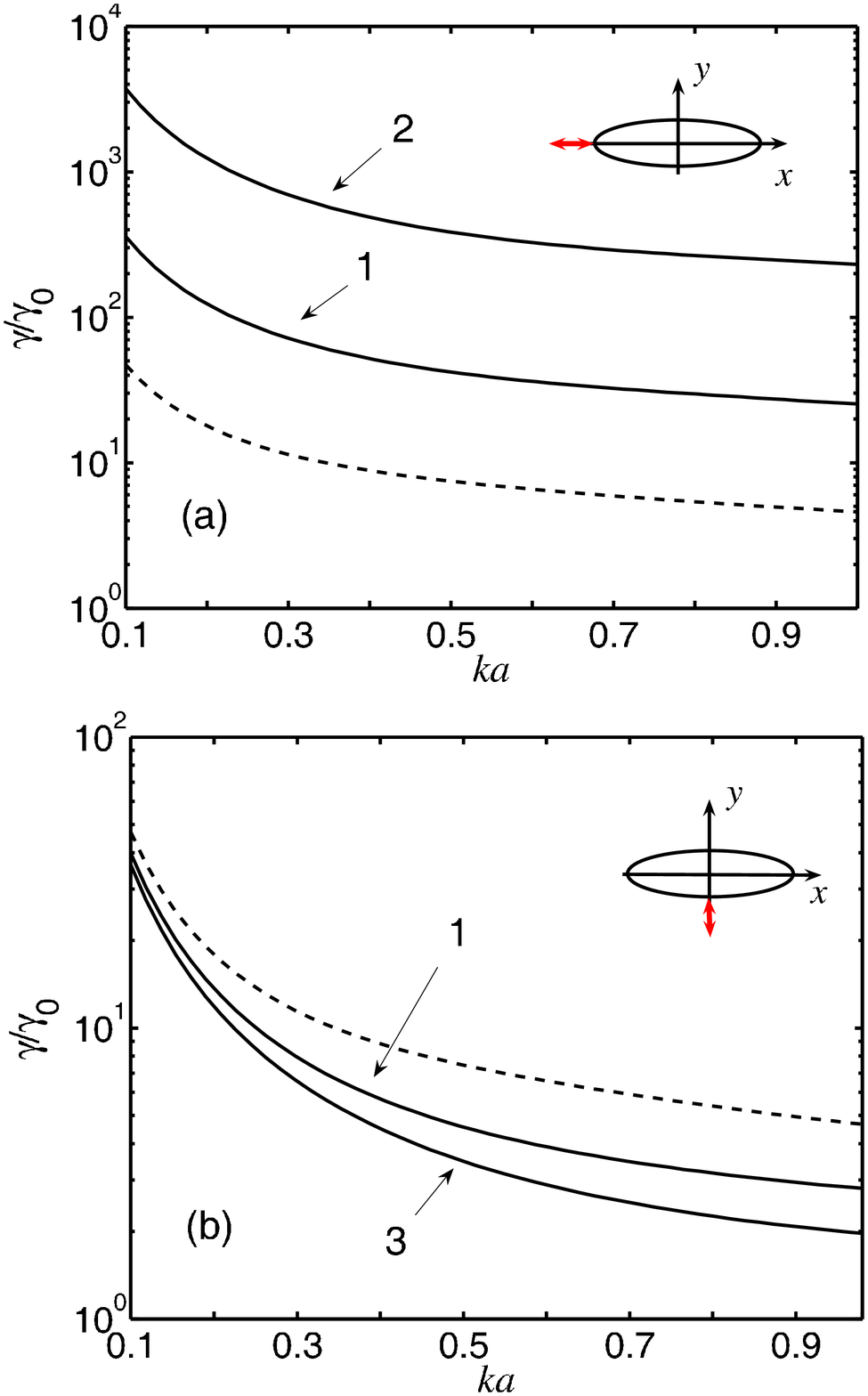}
\caption{}
\end{figure}
\pagebreak

\newpage
\begin{figure}
\centering \includegraphics[height=8cm,angle=0]{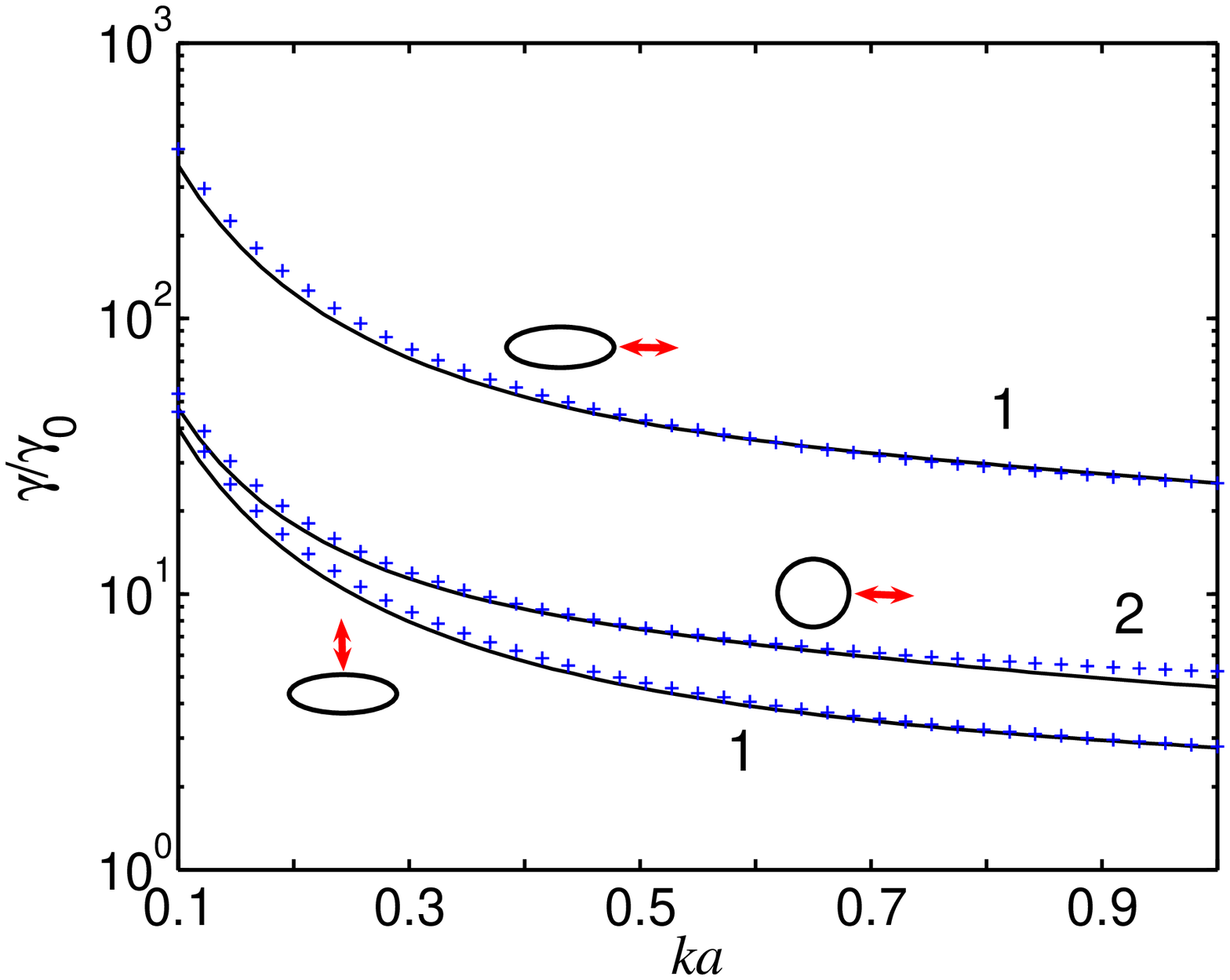}
\caption{}
\end{figure}
\pagebreak

\end{document}